%% file: rv1.tex
\documentclass[twocolumn,trackchanges]{aastex6}
\defcitealias{2016kounkela}{Paper I}

\begin{document}
\title{Kinematics of the Optically Visible YSOs Toward the Orion B Molecular Cloud}

\author{Marina Kounkel\altaffilmark{1}, Lee Hartmann\altaffilmark{1}, Mario Mateo\altaffilmark{1}, John I. Bailey, III.\altaffilmark{2}}
\altaffiltext{1}{Department of Astronomy, University of Michigan, 1085 S. University st., Ann Arbor, MI
48109, USA}
\altaffiltext{2}{Leiden Observatory, Leiden University, P.O. Box 9513, 2300-RA Leiden, The Netherlands}
\email{mkounkel@umich.edu}

\begin{abstract}
We present results from high resolution optical spectra toward 66 young stars in the Orion B molecular cloud to study their kinematics and other properties.  Observations of the H$\alpha$ and Li I 6707 \AA lines are used to check membership and accretion properties. While the stellar radial velocities of in NGC 2068 and L1622 show good agreement with that of the molecular gas, many of the stars in NGC 2024 show a considerable offset. This could be a signature of either expansion of the cluster, high degree of the ejection of the stars from the cluster through the dynamical interaction, or the
acceleration of the gas due to stellar feedback.
\end{abstract}


\section{Introduction}

Most stars begin their lives in clusters \citep[e.g.][]{2003lada}. Therefore, in order to understand the conditions which lead to star formation, it is imperative to understand the dynamical state of the young clusters, with ages less than a crossing time, where the initial conditions have not yet been erased through the dynamical interactions between members. In particular, an important question that is yet to be answered is whether the cluster form quickly on a free-fall time scale \citep[e.g.][]{2007elmegreen,2007hartmann,2015kuznetsova}, or whether the clouds are initially supported by turbulence preventing a rapid collapse \citep[e.g.][]{2006tan,2012hennebelle}.

Several kinematic studies have been conducted in other nearby massive clusters, such as the ONC and NGC 2264 \citep{2006furesz, 2008furesz, 2009tobin, 2015tobin, 2016dario, 2016kounkela}, analyzing the radial velocity (RV) of the young stellar objects (YSOs) within them. These observations revealed that these clusters are not dynamically relaxed and that they show a considerable RV substructure. In particular, they showed that, while typically the RV of the stars are similar to the kinematics of the gas from which they have formed, a large number of stars in both of these clusters are preferentially blueshifted relative to the gas; this blueshifted population is not compensated by an equal number of redshifted sources. In some extreme cases \citep[e.g., toward the Cone Nebula,][hereafter, Paper I]{2016kounkela}, the gas and the stars appear to be entirely decoupled from each other. Some explanations have been proposed to explain the blueshifted population; however, so far there is no conclusive answer.

The Orion B molecular cloud contains several clusters with an ongoing star-formation, such as NGC 2023/2024, 2068/2071, and L1622. These nearby clusters \citep[390---420 pc,][]{2017kounkel} are young, with ages of $<2$ Myr \citep[e.g.][]{2006Levine, 2008Flaherty, 2008Kun}. These clusters are greatly affected by the high degree of extinction; only a few stars out of hundreds of known members have optical emission. Previously, \citet[][hereafter, FM08]{2008Flaherty} were able to obtain high-resolution optical spectra to measure RV to 32 stars in NGC 2068. To our knowledge, no currently published surveys obtained stellar RVs in the other Orion B regions. In the future, the IN-SYNC survey will present high-resolution near infrared spectra taken with APOGEE for the stars in the cloud; this is the extension of the observations of the Orion A molecular cloud \citep{2016dario,2017dario}. While infrared spectra enable the study of many highly-extincted stars that the present optical investigation cannot reach, observations of H$\alpha$ and Li I are important for checking membership and addressing accretion properties.

In this paper we present high-resolution spectral observations of optically emitting stars in the clusters associated with the Orion B molecular cloud. In Section \ref{sec:data} we describe our observations and define the membership sample. In Section \ref{sec:class} we discuss evolutionary classification of the sources and look at the properties of the H$\alpha$ and Li I lines observed toward them. In Section \ref{sec:vel} we look at the distribution of RVs in the clusters.

\section{Observations and data reduction}\label{sec:data}
We observed a total of 4 fields toward the Orion B with Michigan/Magellan Fiber System \citep[M2FS,][]{m2fs}, a multi-object spectrograph on the Magellan Clay Telescope. These fields included regions toward NGC 2023, 2024, 2068, and L1622 (Table \ref{tab:m2fsfield}). Due to their spatial proximity, we consider NGC 2023 and NGC 2024 together in the analysis presented in this paper. All regions were observed with the H$\alpha$ and Li I filter, simultaneously spanning two orders covering the spectral range of 6525---6750\AA~ with a spectral resolution $R\sim20,000$. A maximum of 128 sources can be were observed in this configuration with the field of view of 29' in diameter. NGC 2068 has also been re-observed the second time with the H$\alpha$ and the Li I filter, as well as Mg I filter, which spans the spectral range of 5100---5210\AA.

\begin{figure}
\epsscale{1}
\plotone{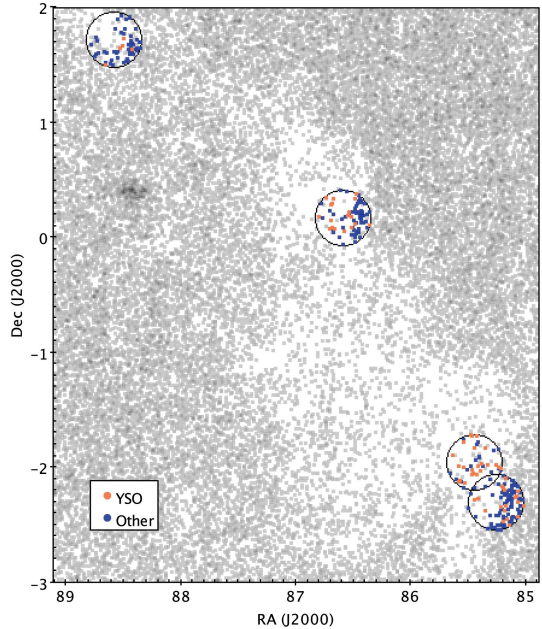}
\caption{The distribution of sources in UCAC4 catalog toward the Orion B (greyscale). The observed fields as well as the sources within them are indicated. Sources with Li I absorption are identified as YSOs. \label{fig:ucac}}
\end{figure}

The observed targets include all the objects found toward the fields from UCAC4 catalog \citep{ucac4}, with the preference for $r$ magnitude brighter than 16.5. Due to extinction, in all four fields there are only 413 sources that are part of the UCAC4 catalog (Figure \ref{fig:ucac}). In contrast, near infrared surveys, such as 2MASS have more than ten times the number of sources in the same part of the sky.

\begin{figure*}
  \centering
		\gridline{\fig{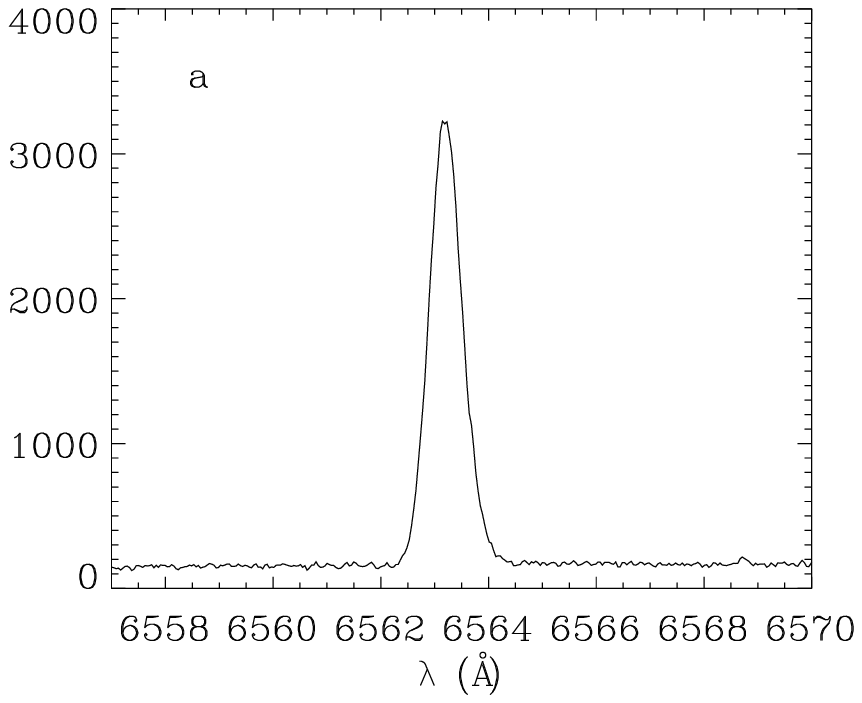}{0.25\textwidth}{}
              \fig{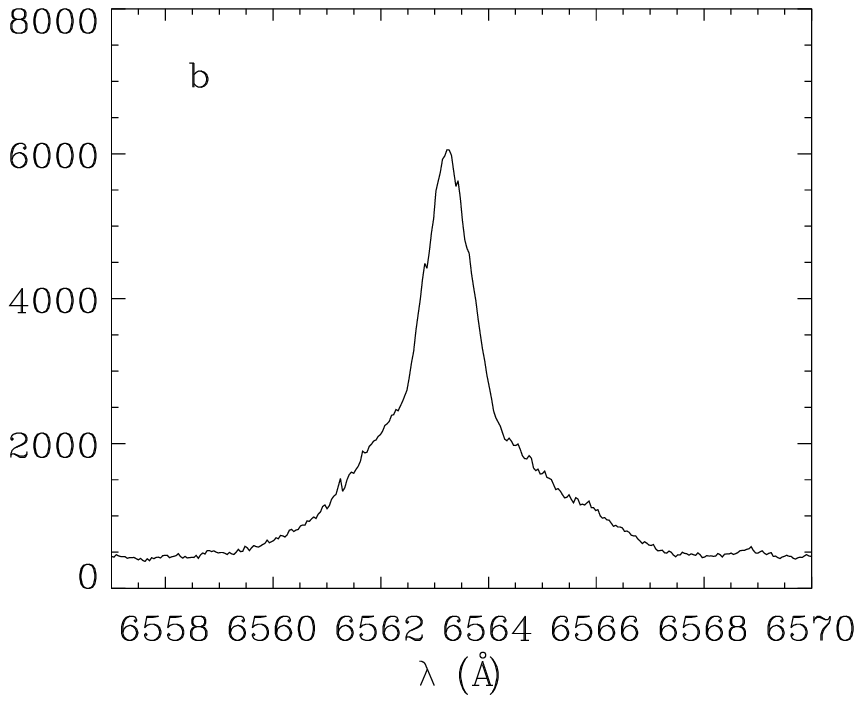}{0.25\textwidth}{}
              \fig{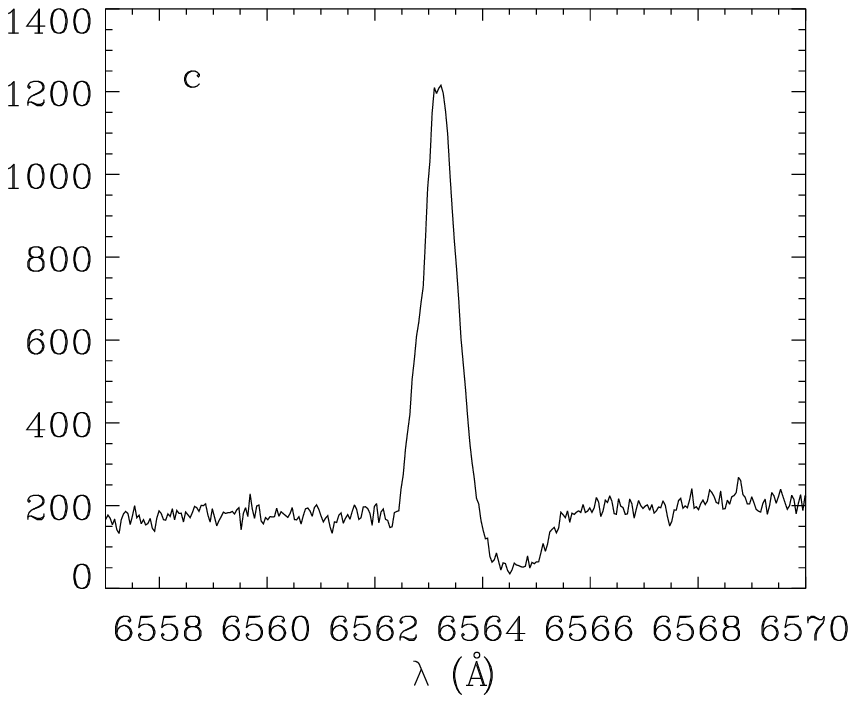}{0.25\textwidth}{}
              \fig{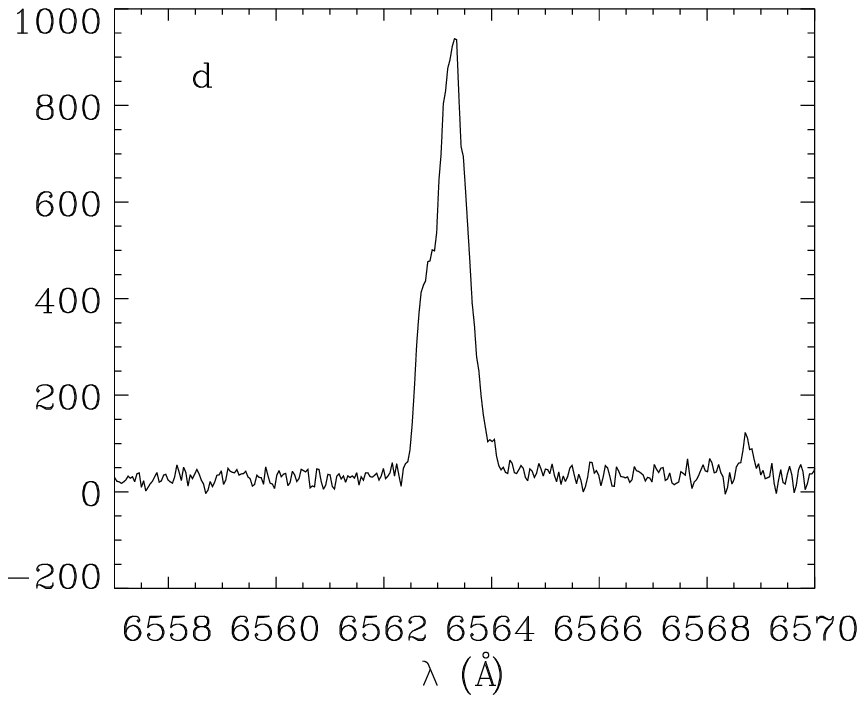}{0.25\textwidth}{}
        }
        \vspace{-0.8cm}
		\gridline{\fig{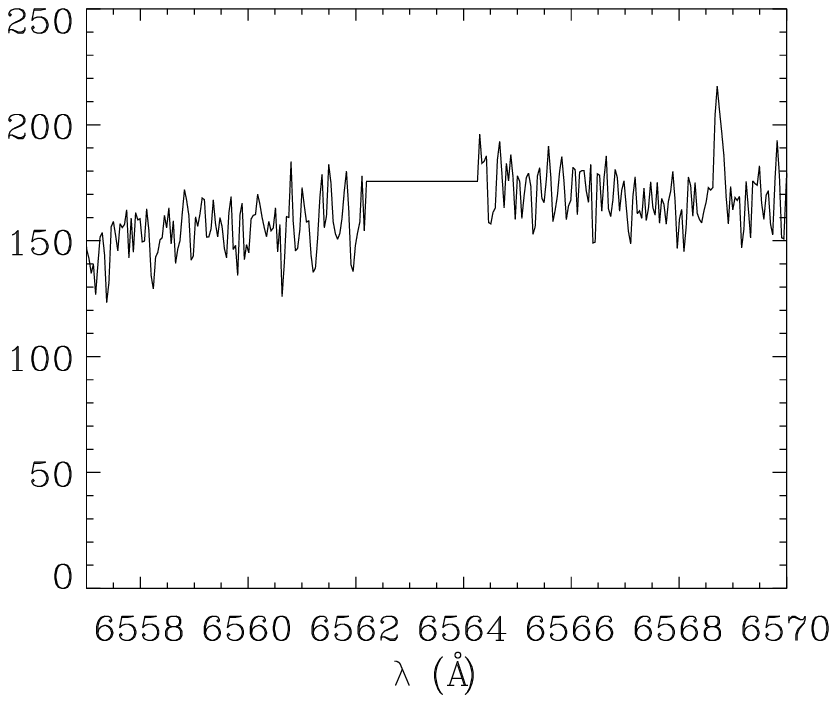}{0.25\textwidth}{}
              \fig{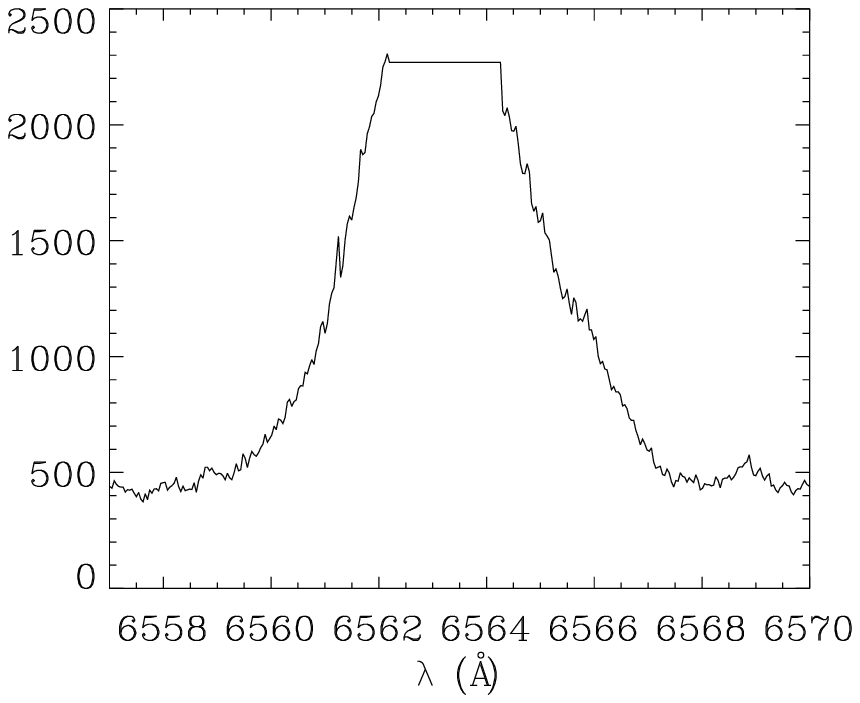}{0.25\textwidth}{}
              \fig{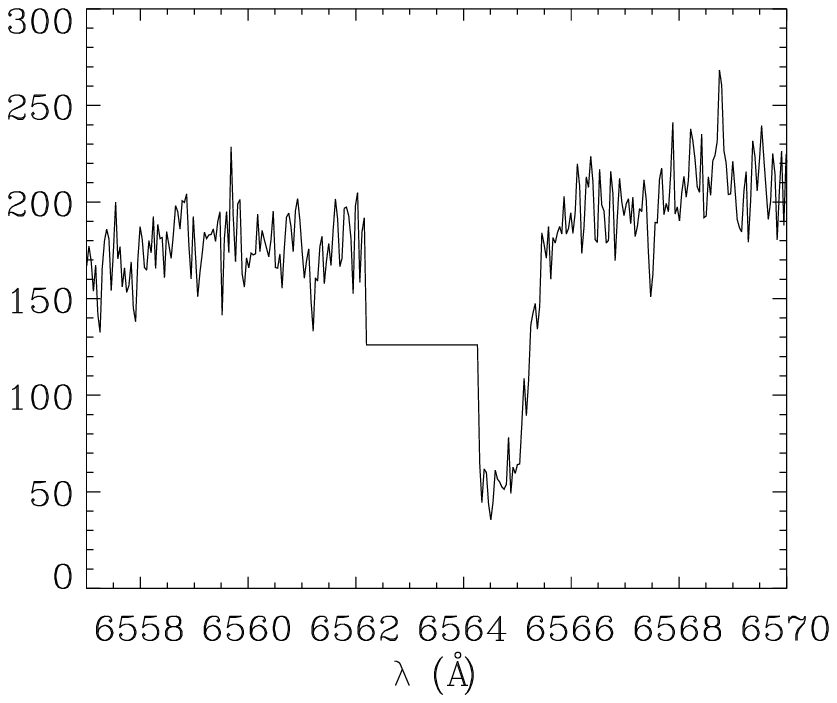}{0.25\textwidth}{}
              \fig{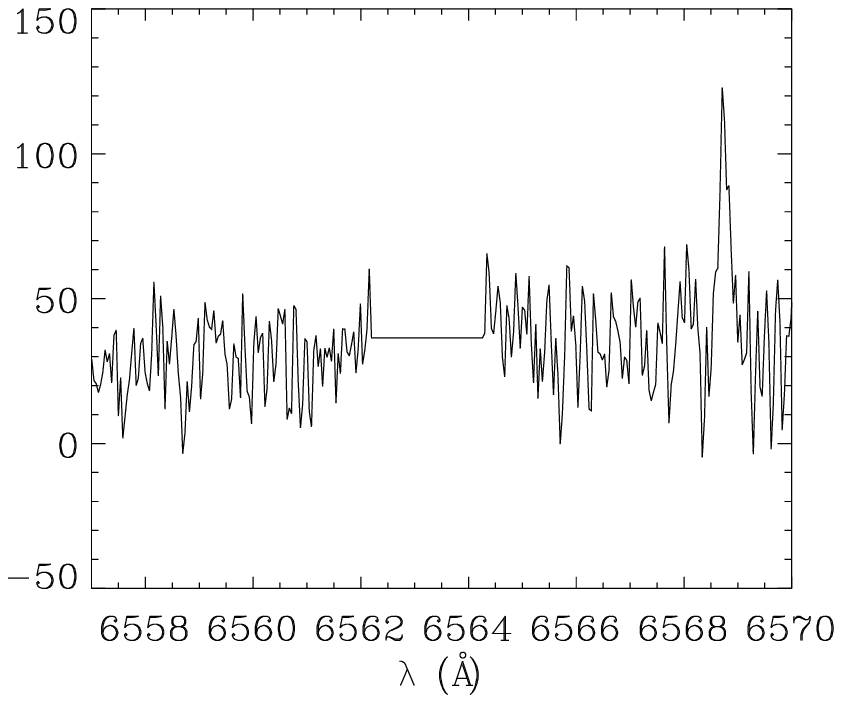}{0.25\textwidth}{}
        }
      \caption{Examples of masking of the H$\alpha$ nebular emission lines. Top row: before masking, bottom row: after masking. a) No residual H$\alpha$ detected coming from the star. b) Strong H$\alpha$ emission. c) Likely H$\alpha$ absorption from a source redshifted relative to the nebular emission. d) No residual H$\alpha$ detected after masking, however, nebular emission line does show some asymmetry in the blue wing which may suggest some H$\alpha$ emission. \label{fig:ha}}
\end{figure*}

The data were first processed by the custom Python code written by J. Bailey, and then reduced using the IRAF pipeline HYDRA. The narrow nebular emissi lines from [S II] (6717 and 6731\AA), [N II] (6549 and 6583\AA), and H$\alpha$ appeared strongly toward many (but not all) sources. While sky spectra offset by a few arcseconds were taken, in many cases, particularly in NGC 2024, it was insufficient to reliably remove this emission which can strongly vary even over small angular distances. These lines were masked out (Figure \ref{fig:ha}). The same was done for the wide Littrow ghosts from the optics at 6600 and 6725\AA~in H$\alpha$ and Li I orders.

The data were cross-correlated with the IRAF package RVSAO \citep{rvsao} relative to the synthetic spectra of \citet{coelho}. We also performed cross-correlation using the synthetic spectra from \citet{munari}, but while overall they produced consistent solutions, \citet{coelho} spectra led to a somewhat better agreement in the solutions between the different simultaneously observed orders for the same stars. The synthetic spectra templates that were used had the effective temperatures ($T_{eff}$) between 3500 and 7000 K in steps of 250 K, the solar metallicity, and surface gravity $\log(g)=3.5$. We obtained the radial velocities (RV) from the cross-correlations, and estimated $T_{eff}$ from the best-matched template for each object. We include only RV measurements where the signal-to-noise $R$ \citep{1979tonry} of the best cross-correlation was $R>6$. The Li I and H$\alpha$ orders were cross-correlated separately; however, both Li I and H$\alpha$ lines were masked out so as not to introduce additional biases due to the relative strength of these lines in comparison to the template spectra, as both lines are significantly stronger in YSOs. It is worth noting that after masking these lines, there appears to be some bias in the best-matched $T_{eff}$ for the template. In particular, there is a strong preference for very low $T_{eff}$ after masking H$\alpha$ line; similarly, Li I masking may result in somewhat higher matched $T_{eff}$. This should not have a strong effect on the RV as the wavelength of the remaining lines is not affected. The best matched $T_{eff}$ between all orders (including Mg) appears to be more consistent if the Li I and H$\alpha$ lines are left unmasked. Therefore, the quoted $T_{eff}$ that we include in Table \ref{tab:li} come from the unmasked cross-correlations.

Generally, the strongest feature in the order containing H$\alpha$ line is H$\alpha$ itself with few other lines present. The Li I order typically provides more reliable RV measurements. If the velocities obtained from each order for the same source differed by less than the uncertainties of each correlation added in quadrature, then the variance-weighted average was calculated for the RV and the uncertainty $\sigma$ of the fit, and the two $R$ values were added in quadrature. If the RVs from the two orders were larger than the expected errors, only the RV from the Li I order was used. RVs from the H$\alpha$ order were only adopted when no significant measurement was possible from the Li I order. The RV measurements from the Mg I order are presented separately.

In addition to the default filtering parameters of the cross-correlation to filter the noise and large scale structure, we used filtering parameters more suited for the rapidly rotating stars, and we recorded the resulting cross-correlation if the resulting $R$ value was greater than in the default case, and the $\sigma$ was not greater by more than 0.05 km s$^{-1}$. The complete description of the data reduction and cross correlation methods is presented in Paper I.

Some observations in the first epoch of NGC 2068 and L1622 have been contaminated by moonlight. The measured RVs of these stars are consistent with the barycentric velocity on the dates of the observations ($\sim -25$ km s$^{-1}$), and the best-matched temperature template consistent with solar. The sources in which this contamination has been identified have been removed from the source list.

The 2015 epoch of observations of NGC 2068 field appears to be systematically blueshifted by 2 km s$^{-1}$ relative to the 2017 epoch. There are not many telluric lines in the wavelength regime covered by the spectra (most apparent lines are found at 6542.313, 6543.907, 6547.705, 6572.086 6574.852\AA), and they are generally relatively weak and not apparent toward most of the sources. However in the sources where it is possible to centroid these lines and it is clear that they are not contaminated by any other nearby lines, we can confirm that 2017 epoch for NGC 2068 field has accurately calibrated wavelength solution and 2015 epoch is the one that is responsible for the offset.

Additionally, a systematic redshift of $\sim$1 km s$^{-1}$ can be observed in the data in Paper I as well in the M2FS observations of H$\alpha$ order and the Li I order, although the latter order is apparent only for non-members. As the Li I line was not masked out previously in members, and the templates do not have as strong Li I absorption, the cross-correlation was caught on this line in the YSOs, producing a slight RV shift, which happened to almost exactly cancel out the systematic offset for the entire field.

There is some evidence that NGC 2024 field is systematically blueshifted by $\sim$2 km s$^{-1}$, and NGC 2023 field is systematically redshifted by $\sim$2 km s$^{-1}$ (see Section \ref{sec:vel} for discussion). This offset is suggested by the telluric lines, although there are only a few sources where centroiding of these lines is possible. We unfortunately cannot confirm it since these fields have not been reobserved and there is a lack of previously published RV measurements of these sources. Nonetheless, we do correct all these offsets from the RVs in the Table \ref{tab:li}.

The causes of these offsets are not clear. It has been observed by \citet{2015Walker} that temperature dependent offsets in the zero point of $\sim$2 km s$^{-1}$ can occur in the M2FS observations; since then a greater care has been taken in the calibration of the instrument. We can rule out the temperature dependence in this set of the observations, as the solutions are consistent between the individual frames. It is possible that the issue is isolated to the specific H$\alpha$/Li I filter set. Nonetheless, while the absolute zero point calibration may be uncertain, the relative RVs within each field should be consistent. 

\begin{figure}
\epsscale{0.8}
\plotone{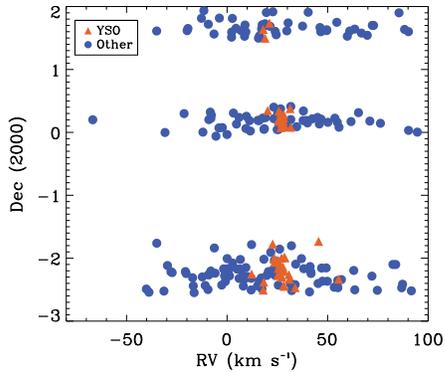}
\caption{Measured RVs of all the sources observed toward the Orion B. \label{fig:allrv}}
\end{figure}

\section{Spectral properties} \label{sec:class}

\begin{figure}
 \centering 
		\gridline{\fig{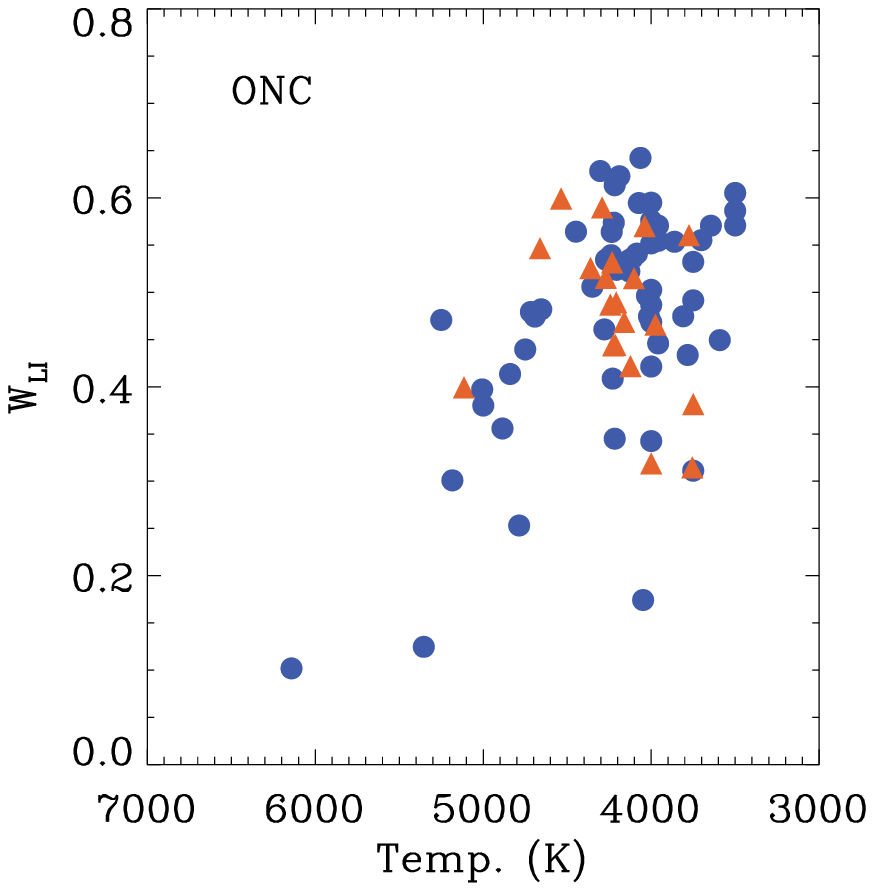}{0.25\textwidth}{}
              \fig{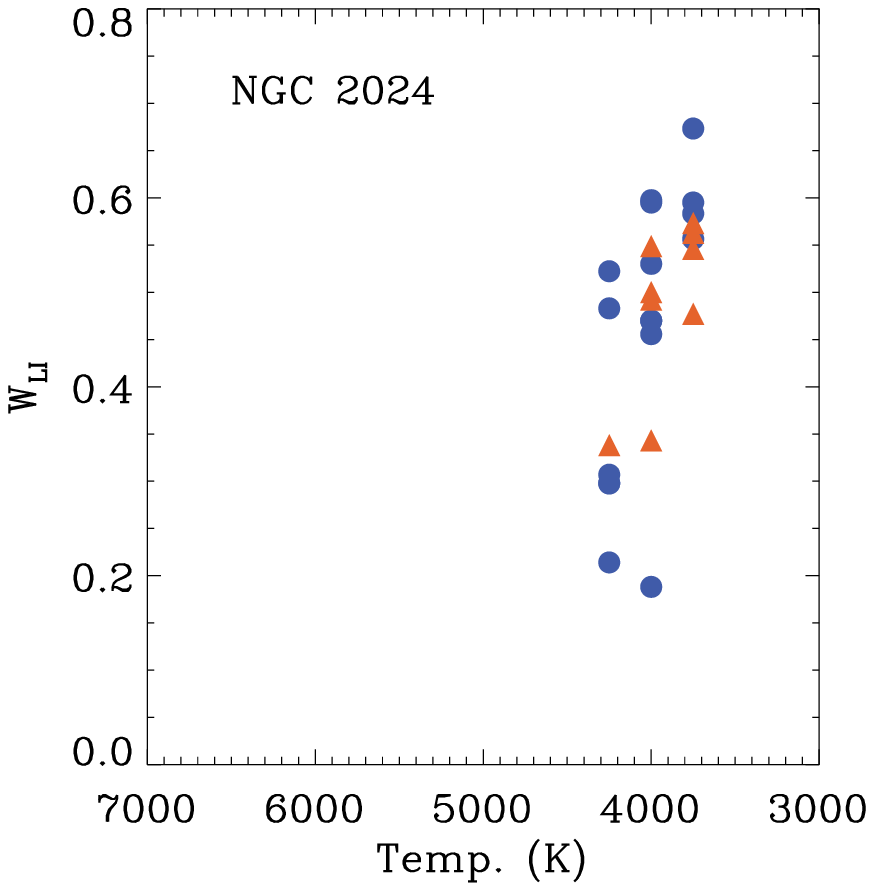}{0.25\textwidth}{}
        }
        \vspace{-0.8cm}
		\gridline{\fig{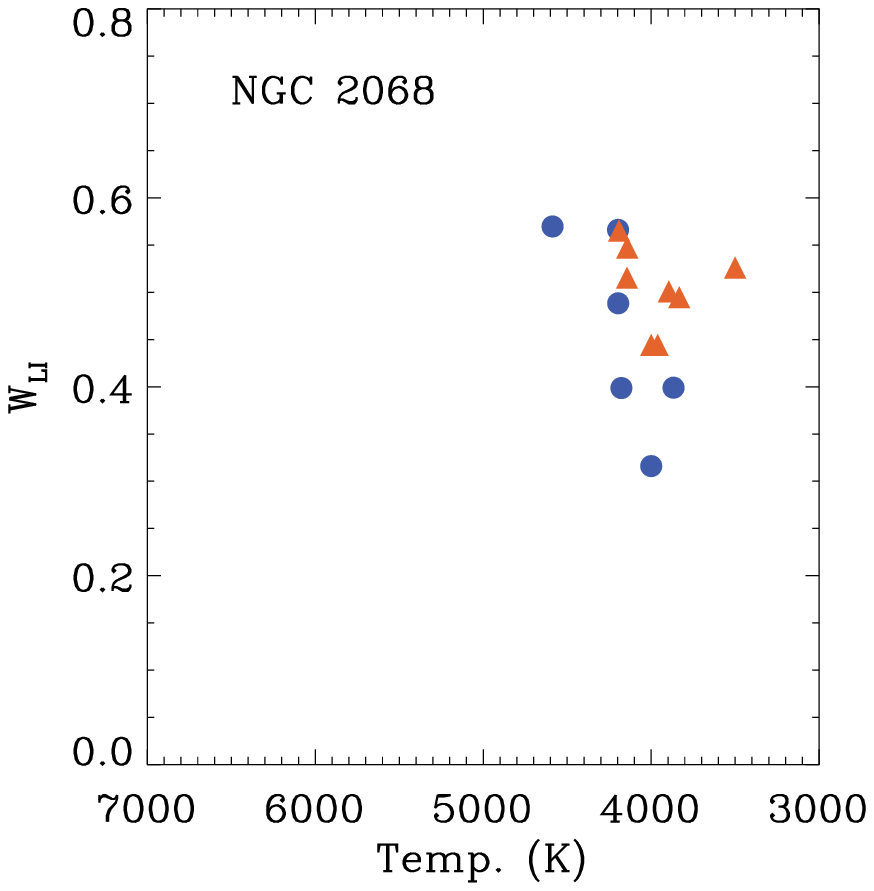}{0.25\textwidth}{}
              \fig{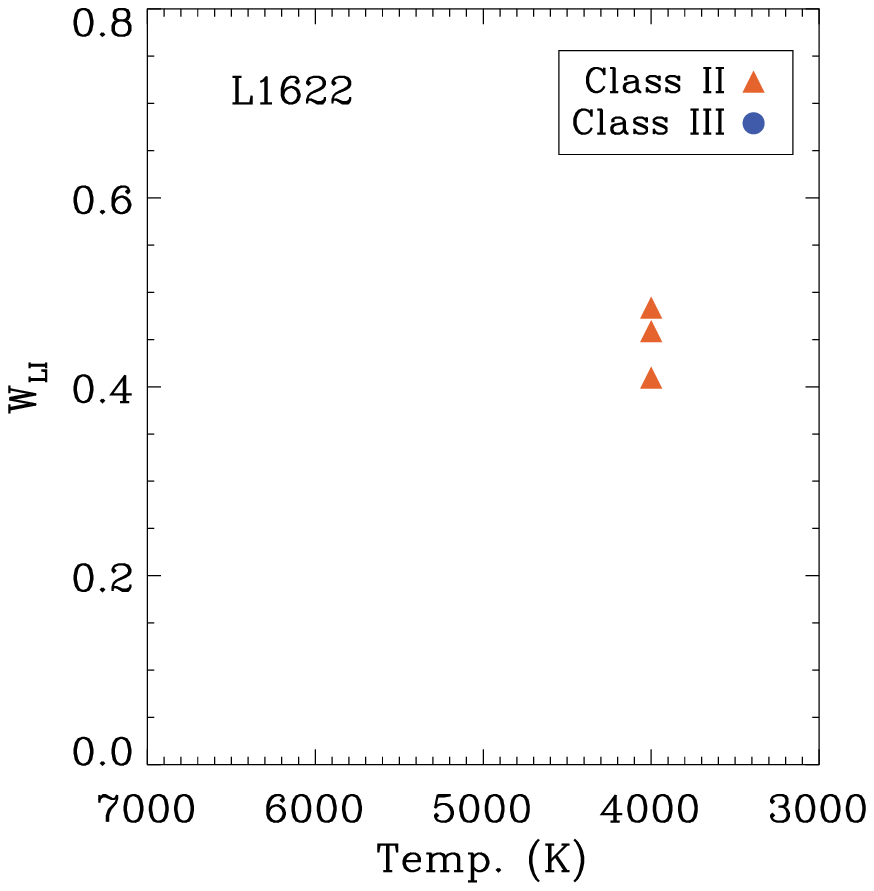}{0.25\textwidth}{}
        }
 \caption{Equivalent width of Li I as a function of the best-matched cross-correlation template temperature (with R$>$6) in the four regions. Data for the ONC is taken from Paper I\label{fig:li}}
\end{figure}

\begin{figure}
\epsscale{1}
\plottwo{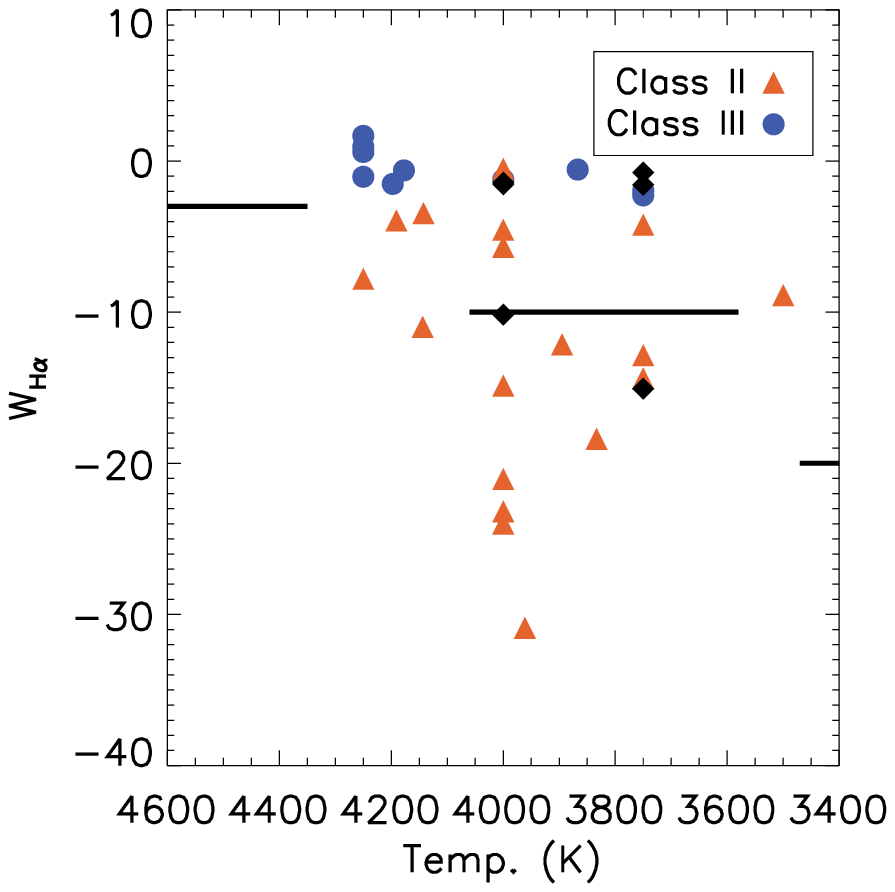}{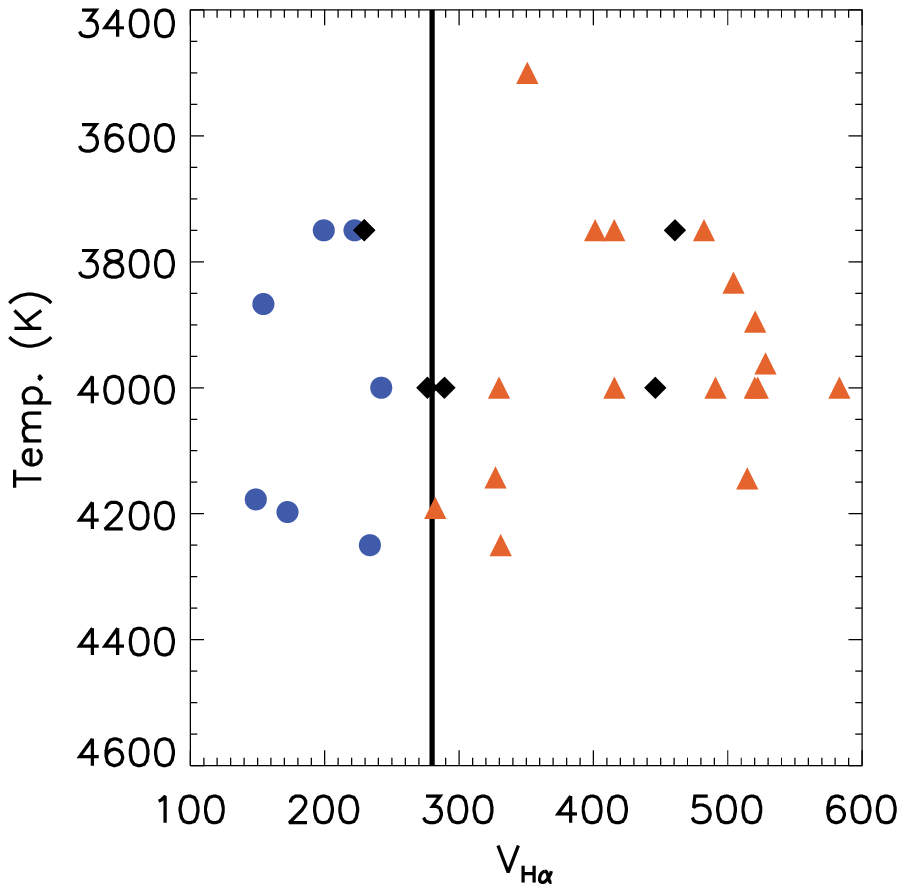}
\caption{Comparison of classification based on the \textit{Spitzer} photometry from \citet{2012megeath} and the accretion signatures from H$\alpha$. Black diamonds are sources not in the \textit{Spitzer} catalog. The black lines show the criteria separating CTTSs and WTTSs from \citet{2003white}. \label{fig:class}}
\end{figure}
\begin{figure}
  \centering
		\gridline{\fig{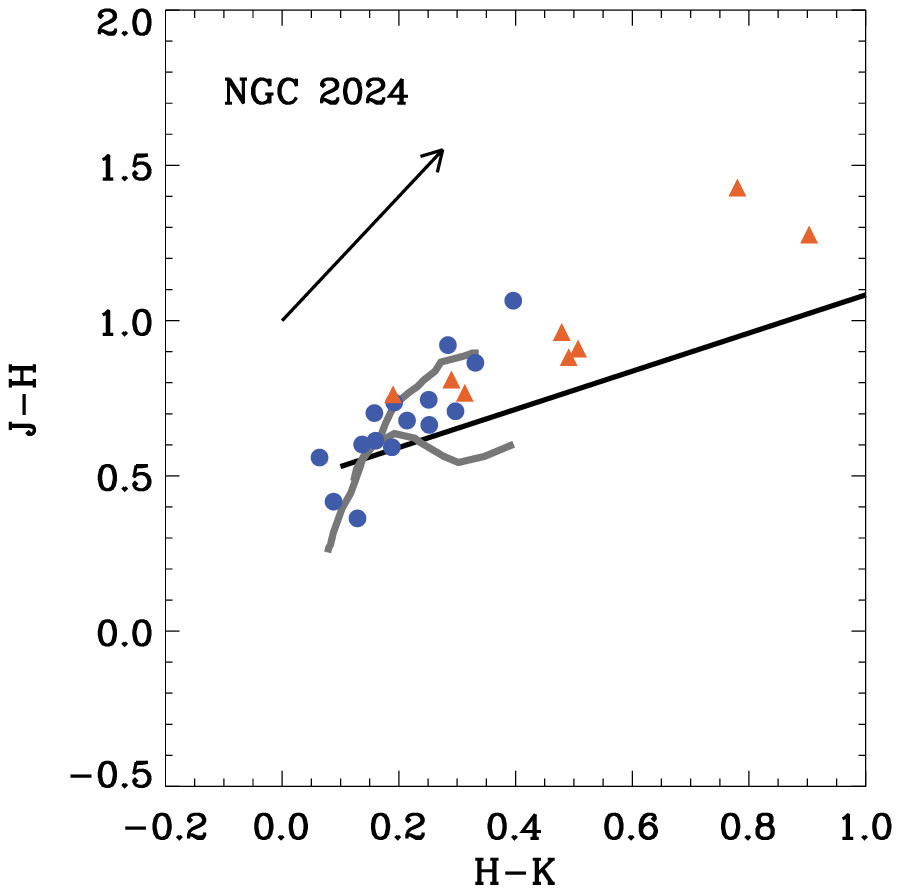}{0.25\textwidth}{}
		      \fig{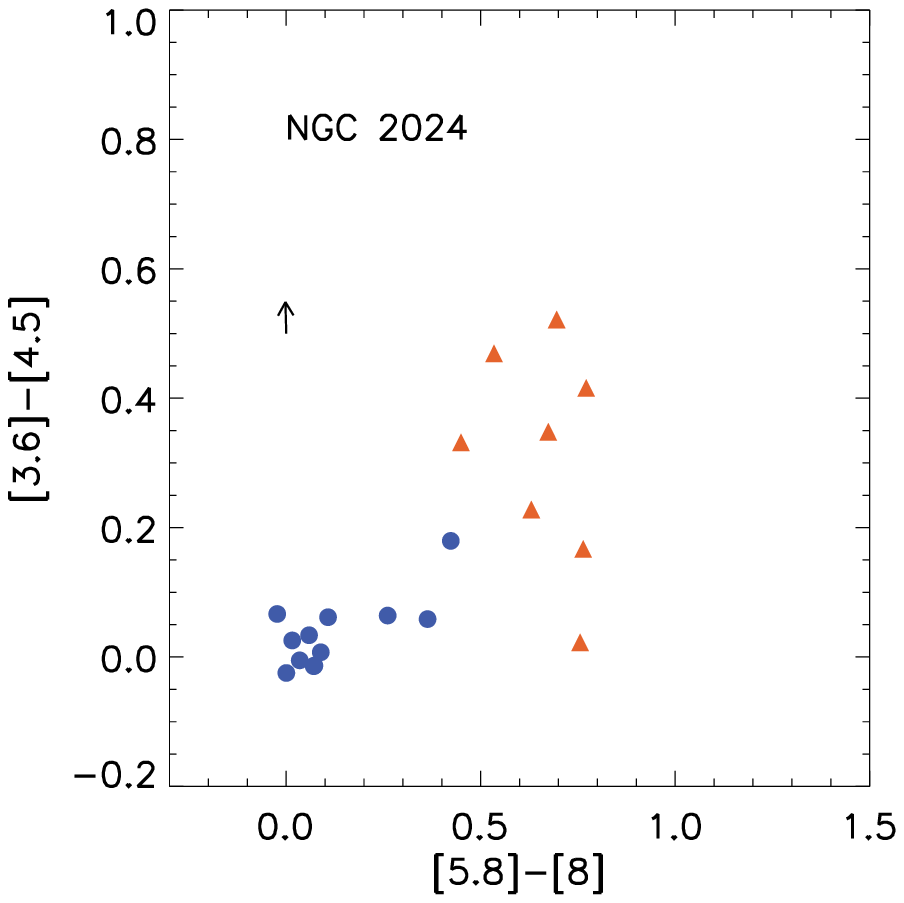}{0.25\textwidth}{}
        }
        \vspace{-0.8cm}
		\gridline{\fig{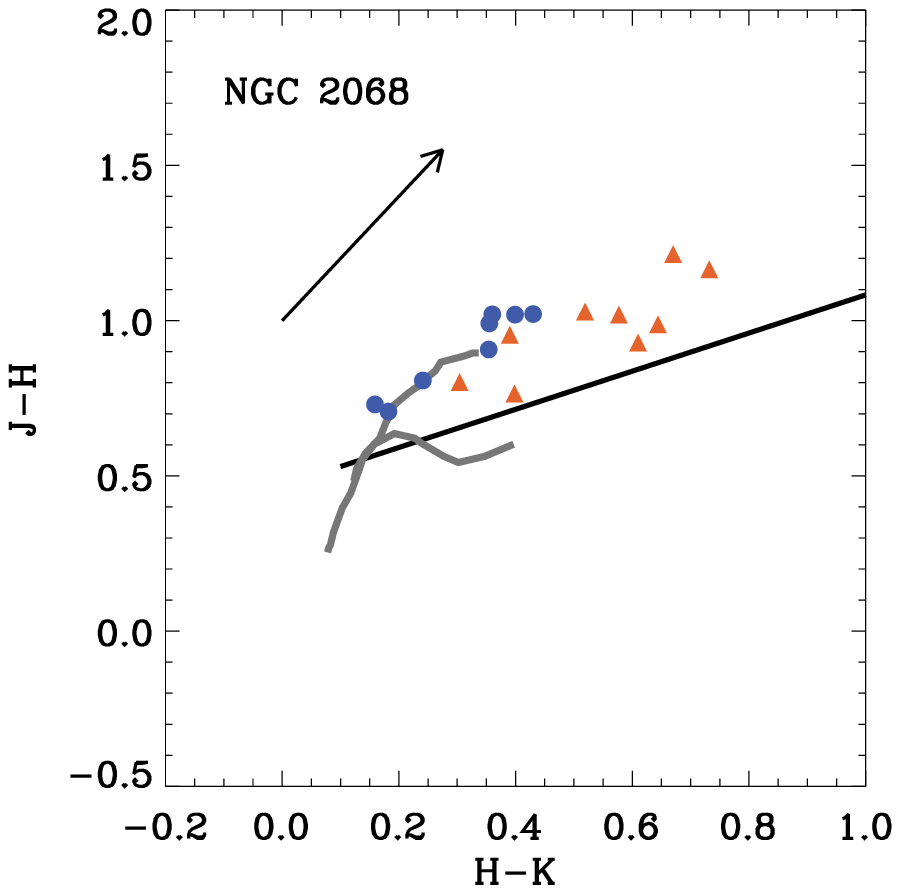}{0.25\textwidth}{}
              \fig{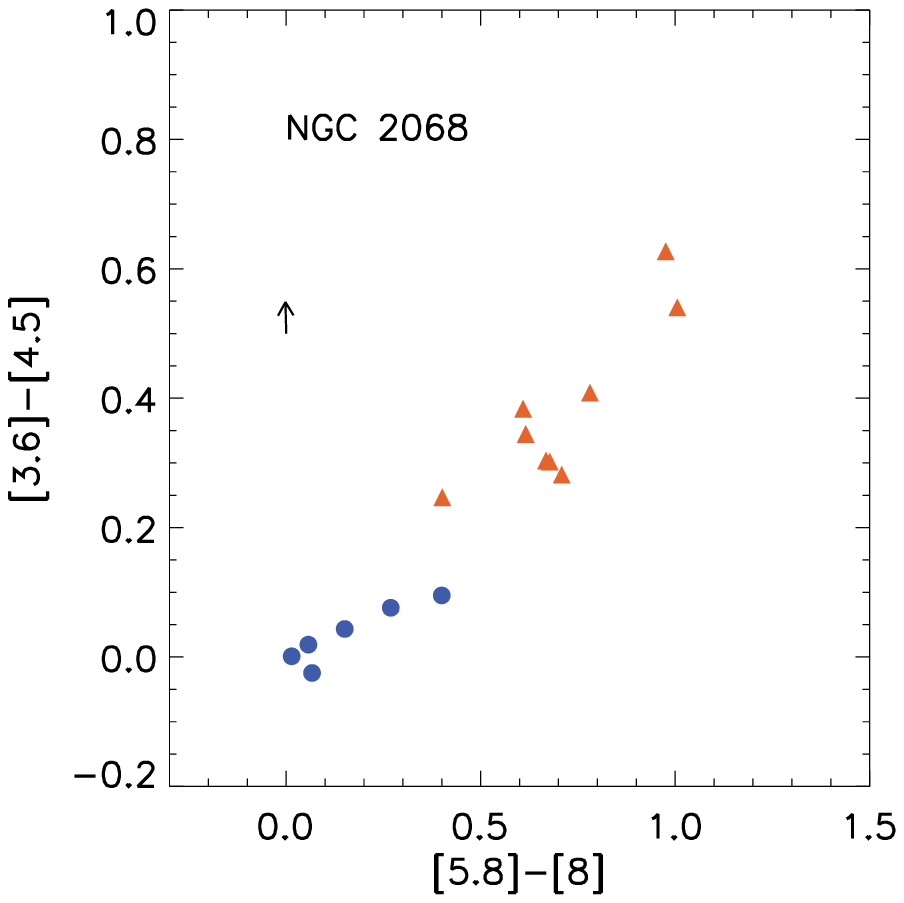}{0.25\textwidth}{}
        }
        \vspace{-0.8cm}
        	\gridline{\fig{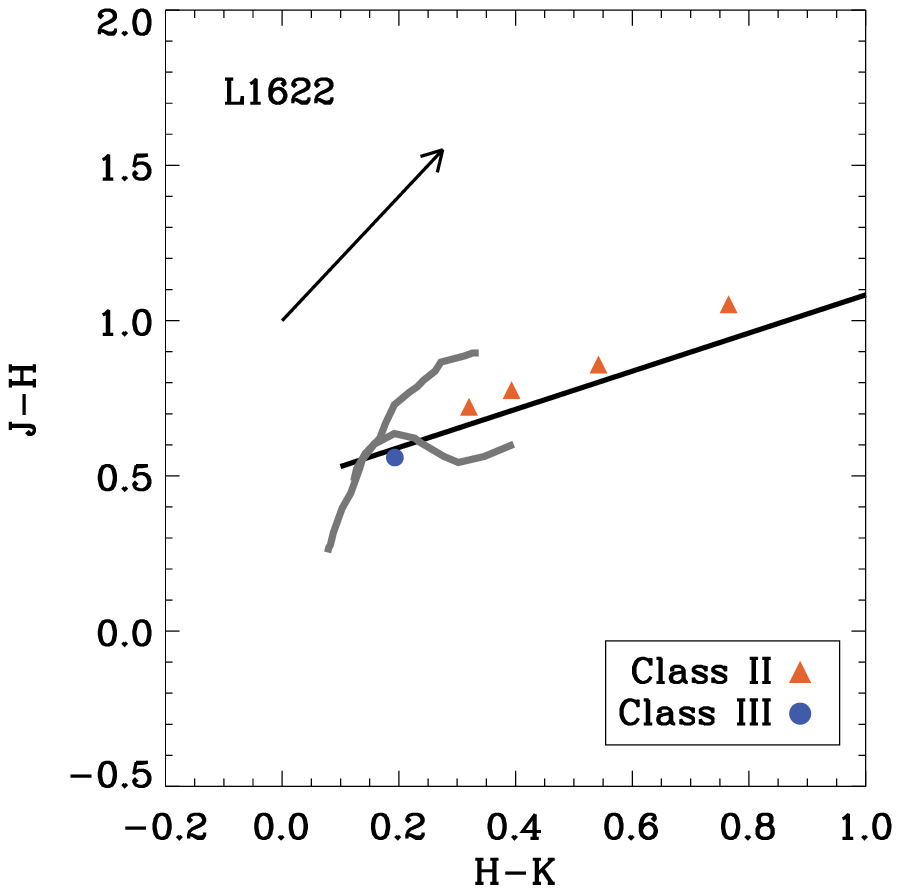}{0.25\textwidth}{}
              \fig{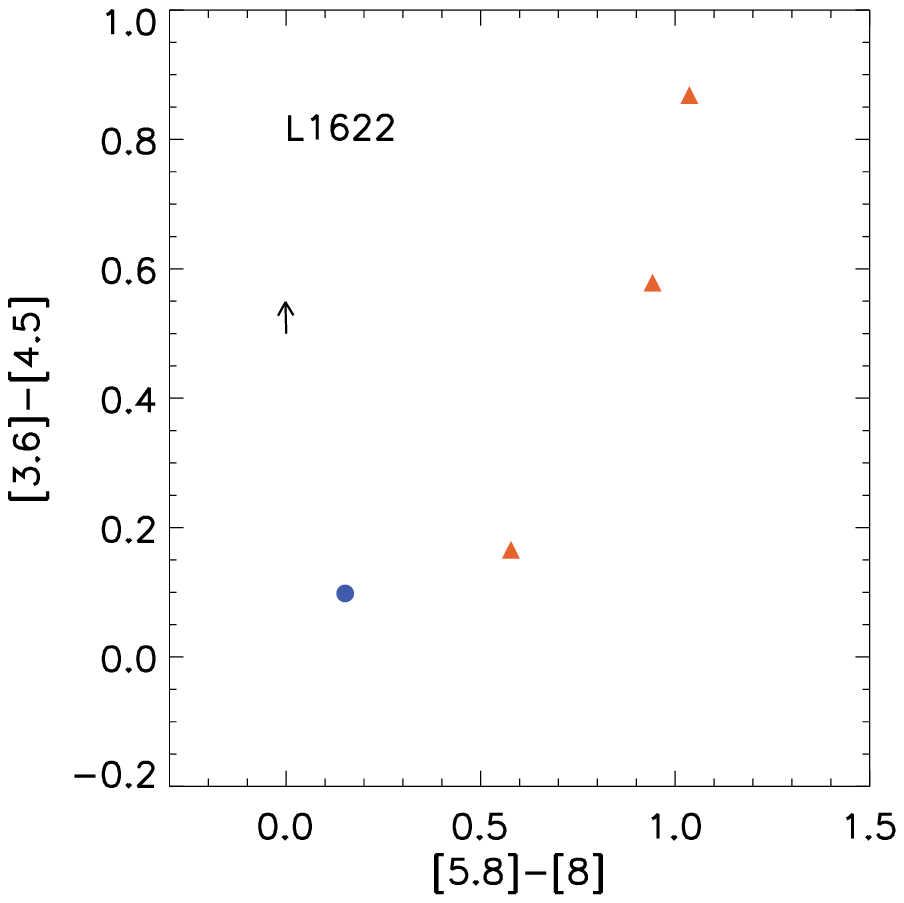}{0.25\textwidth}{}
        }
      \caption{Color-color diagrams for sources with Li I, using \textit{Spitzer} and 2MASS photometry, identified according to their evolutionary classification. Arrows show the reddening vector of 0.5 $A_K$, from \cite{2012megeath}. In the left panels, a black line shows the location of the CTTS locus as identified by \cite{1997meyer}, which corresponds to the intrinsic de-reddened colors of the young stars with disks. A gray lines shows typical colors for pure photospheres \citep{1988bessell}.\label{fig:color}}
\end{figure}

A total of 151 sources have been detected toward NGC 2023/2024, 80 toward NGC 2068, and 59 toward L1622, they are presented in the Table \ref{tab:li}. Figure \ref{fig:allrv} shows a wide spread of radial velocities, much larger than typically observed toward star-forming regions; most of these stars are non-members. We identify members in two ways: strong/broad H$\alpha$ emission due to accretion in the classical T-Tauri stars (CTTS), along with Li I absorption; and Li absorption alone for the weak-lined T-Tauri stars (WTTS). Stars with spectral types earlier than K4 show depleted Li I abundances even during the young age compared to their lower mass counterparts, therefore, uniform membership can be categorized only for stars with $T_{eff}<4500$K (Figure \ref{fig:li}). We report on Li I measurements of the sources with $W_{Li I}>0.1$. We identify 42 sources in NGC 2024 that satisfy this criterion, 18 in NGC 2068, and 6 in L1622; these sources can be confirmed as the members of Orion B. Among the remaining sources, some YSOs with a higher T$_{eff}$ may remain, but most of them are likely to be foreground stars, or possibly background stars if they are not projected directly onto the cloud. CTTSs are identified using the criteria from \citet{2003white}, based on W$_{H\alpha}$ and the velocity width at 10\% maximum of H$\alpha$ ($V_{H\alpha}$, Figure \ref{fig:class}). However, because of the nebular emission contamination, some W$_{H\alpha}$ measurements may not necessarily be reliable. To test this classification, we use classification from \textit{Spitzer} photometry from \citet[][Figure \ref{fig:color}]{2012megeath}. In general, Class II sources (identified by the infrared excess attributed to a dusty disk surrounding them) are expected to be actively accreting CTTSs, and Class III sources (the onese that have colors similar to a naked photosphere) are typically WTTSs, however, the two classifications do not correlate in all cases. Nonetheless, in general, two classifications are comparable with each other. Throughout the text unless stated otherwise we use SED classification with exception of for 7 sources which were not detected with \textit{Spitzer}. These sources were classified based on their H$\alpha$.

Two sources, RV 2670 and RV 2734, are double-line spectroscopic binaries. RV 2670 is a member of NGC 2024. RV 2734 is found in NGC 2068 field; however, it is probably not associated with the cluster. We extract the RVs of the second component from fitting a Gaussian to the cross-correlation function (Table \ref{tab:li}).

We compare the distribution of the $W_{Li I}$ for four regions (including the ONC data from Paper I) in Figure \ref{fig:li}. The typical uncertainty in $W_{Li I}$ of lines with $R>6$ is $\sim$0.05 \AA, although it can increase up to 0.1 \AA~ for noisier spectra. Generally, there does not appear to be any significant differences between CTTSs and WTTSs in terms of the distribution of their $W_{Li I}$. However, accreting stars typically have a large degree of veiling, which may result in underpredicting the true $W_{Li I}$ of these stars. Due to the limited spectral coverage, we cannot correct for veiling in this survey.

Among the non-accreting YSOs (which should not be affected by veiling), there are several sources that appear to be somewhat depleted in Li I relative to the rest. The two most extreme examples are RV 906 and RV 2603 with $W_{Li I}<0.2$. RV 2603 may be an older foreground source, as its velocity is considerably different from the RV of most other identified members of NGC 2024; however, we cannot definitively confirm it. RV 906, on the other hand, was identified as a spectroscopic binary in Paper I, with most measurements kinematically similar to the rest of the ONC. Nonetheless, it is possible that this source is not a member of the ONC, as there are 15 other stars which have no Li I at all that are also found at the rest velocity of the cluster.

A number of other WTTSs have $W_{Li I}<0.4$. Previously, \citet{2005Palla} also discovered 4 stars in the ONC that have shown a similar level of depletion and found their ages to be $\gtrsim$10 Myr, considerably older than members of the ONC or the Orion B. In general, YSOs begin to process Li I at age of 5--10 Myr, and become strongly depleted after 10 Myr \citep[e.g.][]{1998baraffe}. In some rare cases, strong episodic accretion bursts could accelerate the depletion to take place on a much faster rate \citep{2010baraffe,2016baraffe}. However, it is also possible there is a source of emission nearby that would increase the continuum, resulting in a lower $W_{Li I}$. We exclude all the sources found near strong nebulosity or other nearby bright sources identified by examining DSS images of the region. The remaining sources are RV 2550, 2581, 2682, 2725, and 2764. We further attempted to subtract the continuum from the spectra of these sources to test the limits of potential contamination, should it be there. It is notable that for RV 2550, in order to bring $W_{Li I}$ toward a more acceptable range after the continuum subtraction, the flux at the center of Li I line should be near zero.

\begin{figure}
 \centering 
		\gridline{\fig{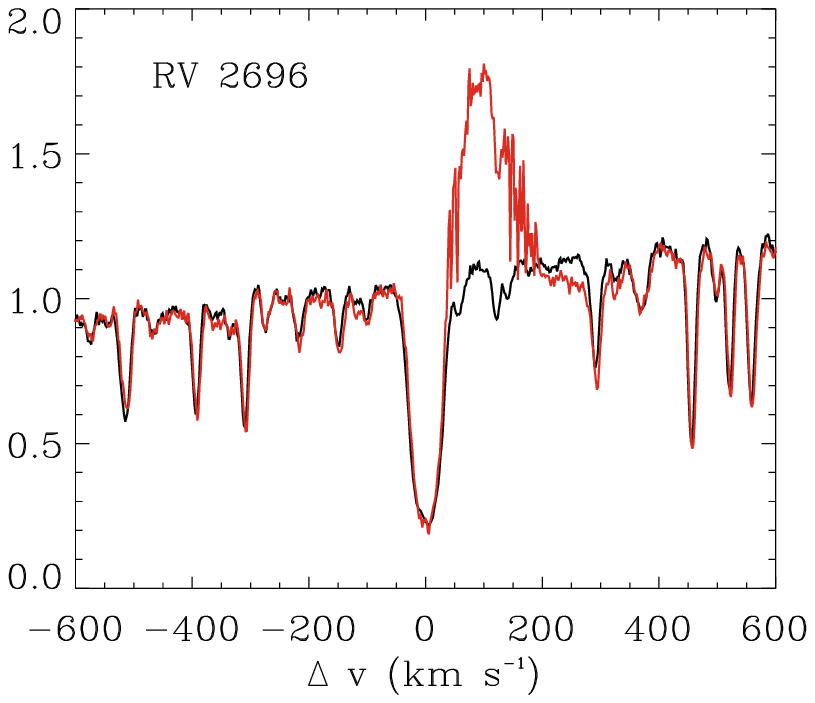}{0.15\textwidth}{}
			  \fig{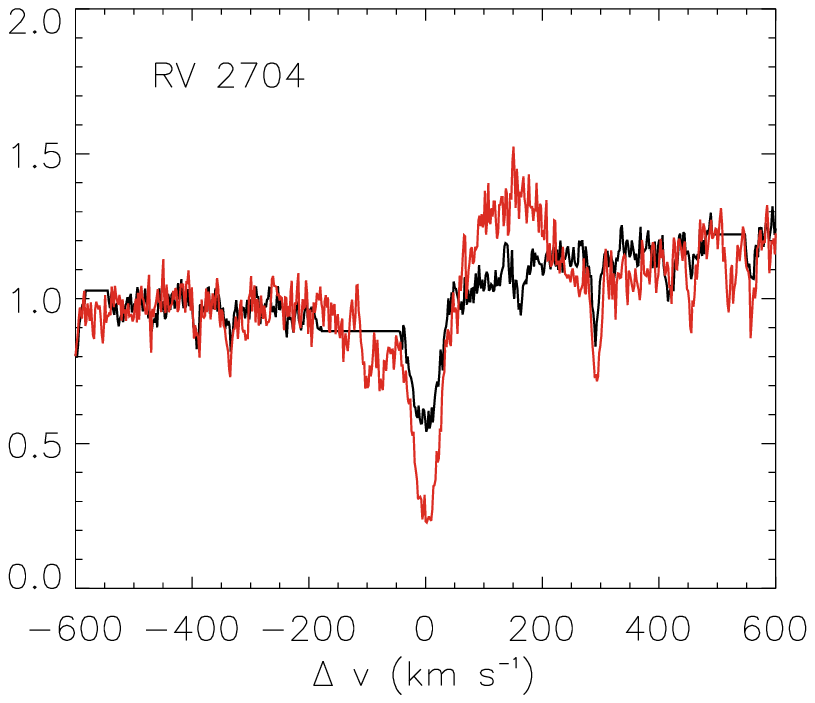}{0.15\textwidth}{}
		      \fig{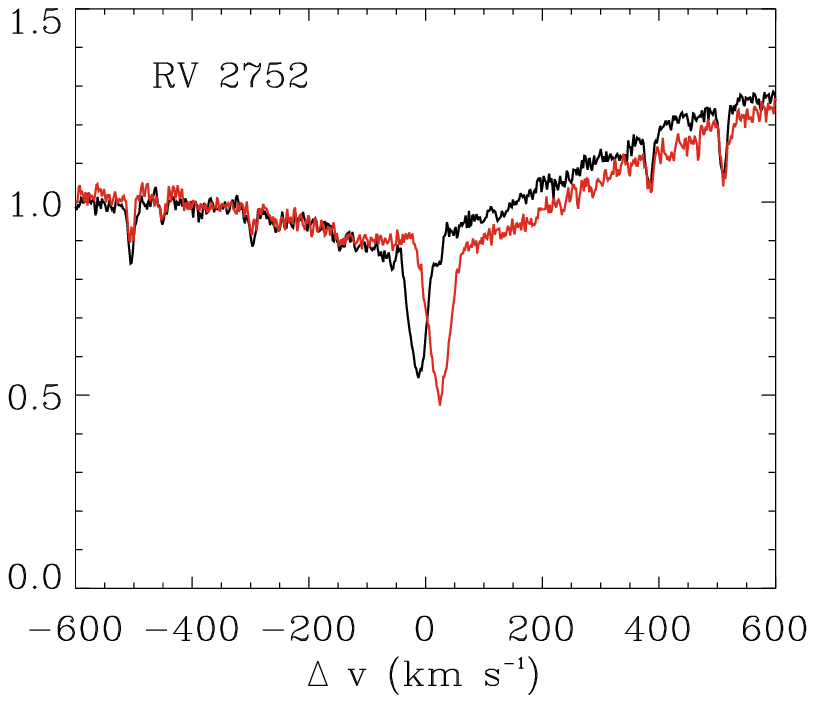}{0.15\textwidth}{}
        }
        \vspace{-0.8cm}
		\gridline{\fig{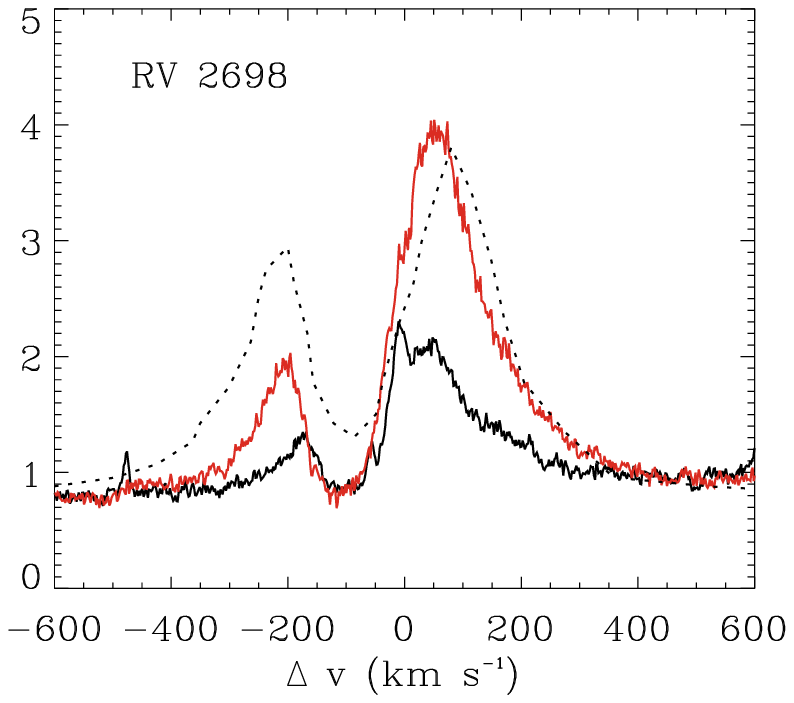}{0.15\textwidth}{}
              \fig{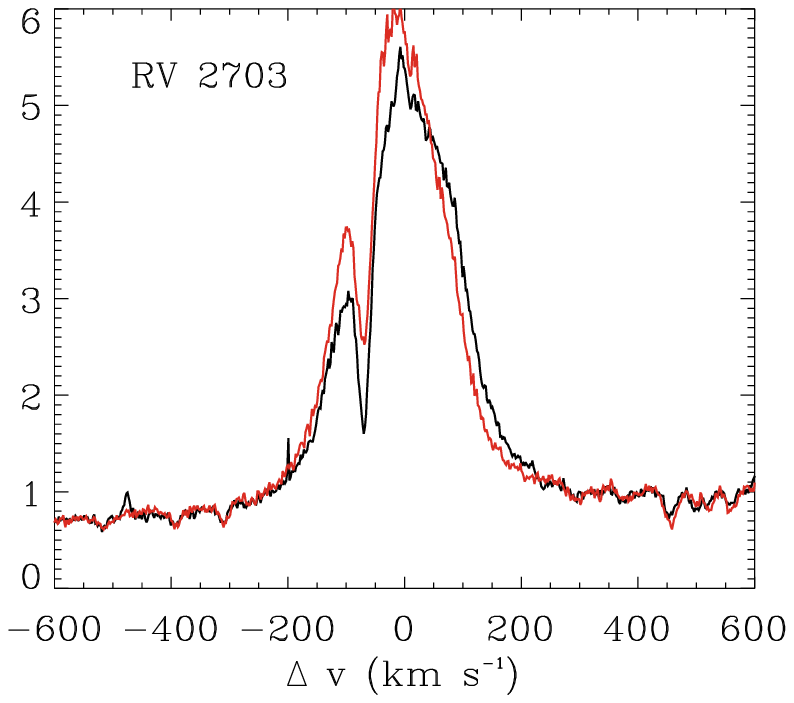}{0.15\textwidth}{}
		      \fig{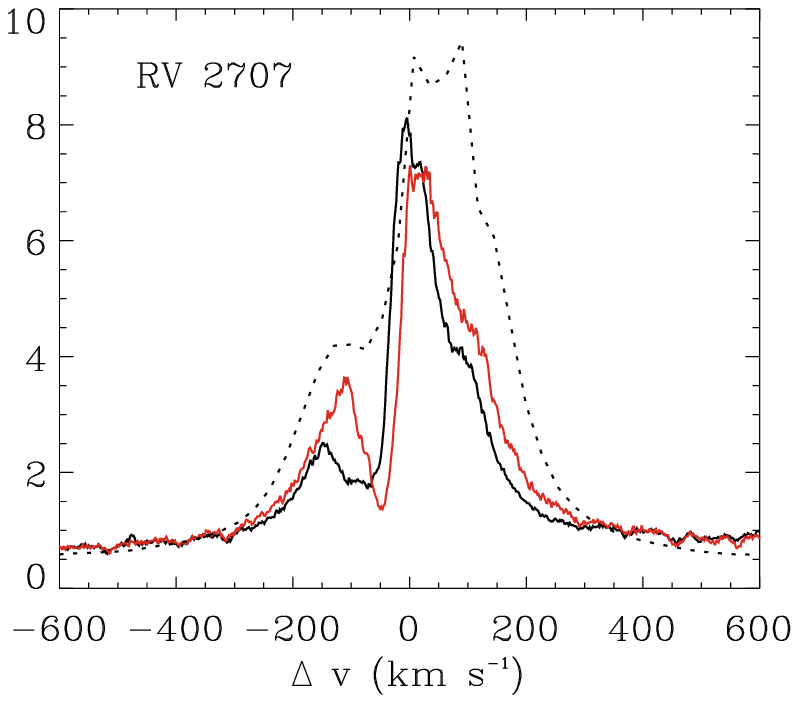}{0.15\textwidth}{}
        }
        \vspace{-0.8cm}
		\gridline{\fig{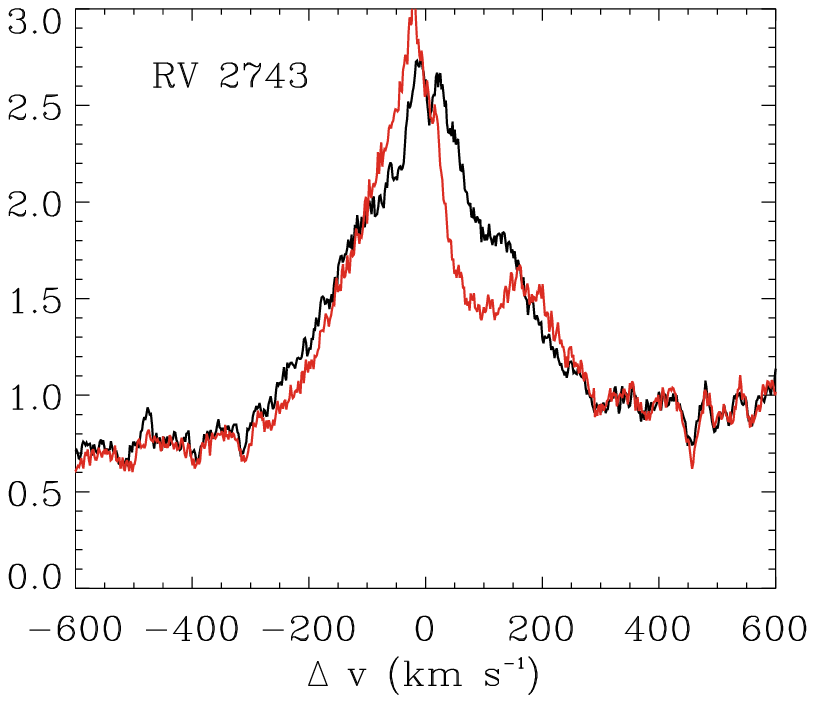}{0.15\textwidth}{}
              \fig{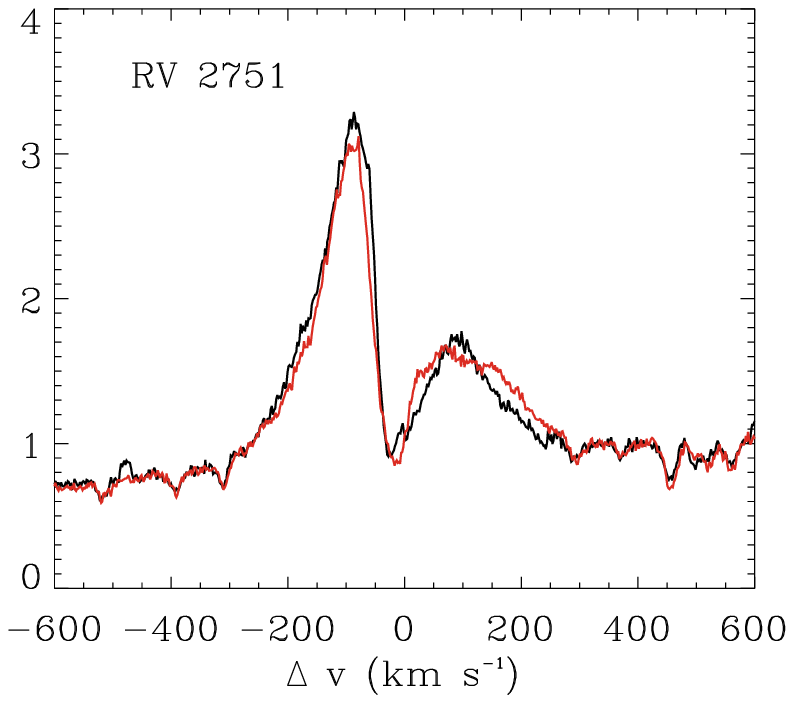}{0.15\textwidth}{}
              \fig{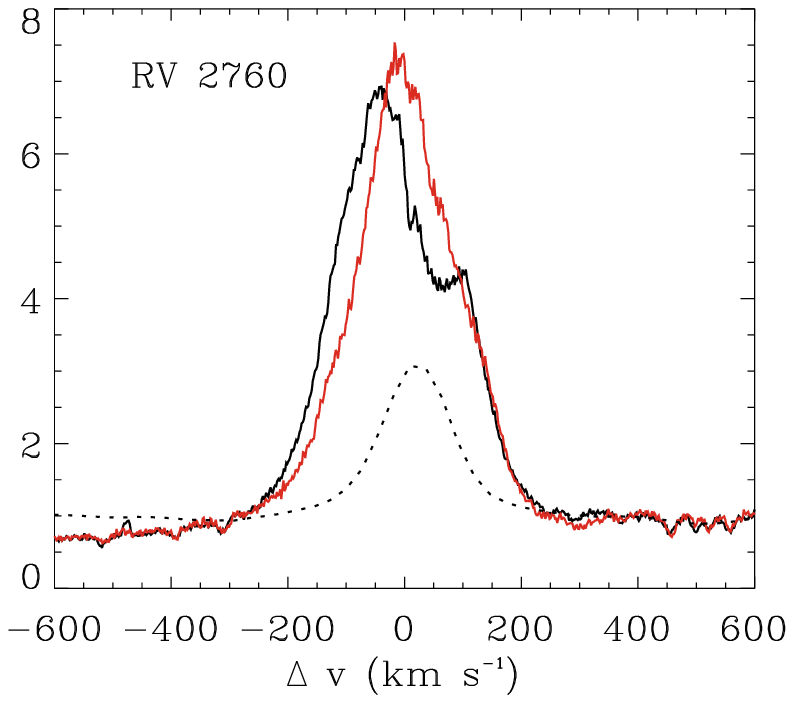}{0.15\textwidth}{}
        }
 \caption{Sources with variable or assymetric H$\alpha$ line. Black curves shows the spectra taken during 2015 epoch, and the red line shows the spectra taken during 2017 epoch. Dashed curves are the spectra from \citet{2009Fang}. \label{fig:varha}}
\end{figure}

One source, RV 2727 does not have any detectable Li I, but it has a clear presence of H$\alpha$ emission (although it is not strong enough to be considered a CTTS by any criteria, although its $V_{H\alpha}$ is close to the threshold, nor does this source have any IR excess). 

Many of the strongly accreting sources do have an assymetric or variable H$\alpha$ line profile (Figure \ref{fig:varha}). These sources are most apparent in NGC 2068 cluster; in L1622 there are too few members observed, and in NGC 2024 the nebular emission is too strong to conclusively analyze the H$\alpha$ line profile.

RV 2696 is not considered to be a member of NGC 2068 since it doesn't have a strong Li I absorption ($W_{Li I}\sim0.05$), although it does have a T$_{eff}\sim5,000$K, at which Li I may become depleted. This source does have RV similar to the RVs of the members of the cluster, however. In the second epoch of observations of this source, it exhibited an emission redward of its H$\alpha$ line that was not present in the first epoch. A similar profile is seen toward RV 2704.

RV 2752 also has very weak Li I line, and therefore is not considered as a member. Its spectrum is very poorly correlated in all observations. This source does have a very broad H$\alpha$ absorption line, it appears to be an early-type star. However, there is a narrow absorption superimposed onto the broad component. The wavelength of this narrow line appears to shift by almost 40 km s$^{-1}$. It is notable that the other lines remain unaffected. It is possible that this is a spectroscopic binary with a late-type companion that is responsible both for Li I absorption and variable H$\alpha$, however, it is difficult to confirm it.

The remaining sources we examine are all CTTSs. The H$\alpha$ emission of RV 2698 has increased in strength by a factor of 3 relative to the continuum between 2015 and 2017. It has an assymetric line profile with two peaks, where the stronger peak is somewhat redshifted relative to the rest velocity of the source, and the dip in the flux is somewhat blueshifted, similar to a P Cygni line profile, although the absorption does not fall significantly below the continuum level. This source has been previously observed by \citet{2009Fang}; the the strength of the primary peak was the same as it is in 2017, although the secondary peak was significantly stronger. Similar profile is seen in RV 2703 and RV 2707, although their absorption component is weaker. The size of the absorption in RV 2707 appears to evolve, becoming narrower over time, and this source has an additional bump in the red part of the line profile. This bump had the same strength as the primary peak in the observations by \citet{2009Fang}.

RV 2743, 2751, and 2760 also have dual peaks, although the absorption is redshifted relative to the center of the line, which is indicative of the infalling gas. RV 2743 shows this only in 2017 epoch of observations, but not in 2015. On the other hand, for RV 2760, this infall is present only in 2015 epoch. Since \citet{2009Fang} observations, the intensity of emission has tripled.

Among the sources identified as members in this survey we measure the accretion ratio defined as [CTTS]/[CTTS+WTTS] ratio to be 0.66$\pm$0.33 (5/6) in L1622, 0.55$\pm$0.18 (10/18) in NGC 2068, and 0.38$\pm$0.09 (16/42) in NGC 2024. The optical regime is somewhat more biased against CTTSs as they may be more reddened and we are sensitive only to the sources with a very low extinction (Figure \ref{fig:color}). Moreover, cluster cores usually contain younger population of YSOs than those found in the halo \citep[e.g.][]{2014getman}, and in Orion B we cannot observe any YSOs embedded in the clouds. Therefore, these values are probably the lower limits of the true disk fractions of these clusters.

\section{Velocity Structure}\label{sec:vel}

We look at the velocity distribution of the confirmed members of the Orion B with detected Li I absorption with the $R>$6. There are 31 such sources in NGC 2024, 13 in NGC 2068, and 4 in L1622. Similarly to Paper I, we compare this velocity distribution to that of the $^{13}$CO gas after converting the measured RV to the local standard of rest (lsr). As the young stars have formed from the molecular gas which is traced by $^{13}$CO, a natural assumption is that both stellar and gas velocity distribution should be comparable to each other.

\begin{figure}
\epsscale{0.6}
\plotone{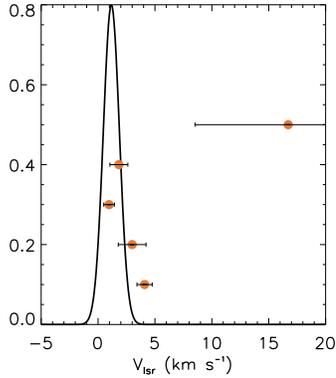}
\caption{Velocity distribution of members toward L1622. Black line shows the distribution of the $^{13}$CO gas described by \citet{2008Kun}. Sources are arbitrarily scaled along the y-axis to maximize readibility. \label{fig:l1622}}
\end{figure}

It does appear to be the case in L1622. In this region, v$_{lsr}$ of the $^{13}$CO gas with the peak 1.17 km s$^{-1}$ and line width at half maximum of 1.66 km s$^{-1}$ \citep{2008Kun}. While the sample of members of L1622 that we detected in this survey is extremely small (only 5 sources, only 3 of which have confident RV measurements with R$>$6), $v_{lsr}$ of all but one members of L1622 ranges between 0.9 to 4.1 km s$^{-1}$ (Figure \ref{fig:l1622}). The RV of the single outlying source is highly uncertain, has low $R$ value. It is possible that this source may be affected by the multiplicity. Unfortunately, only a single epoch of measurements is available to confirm it. It is possible that there may be a systematic offset for the entire field, but it unlikely to significantly change the correlation between RVs of the stars and the molecular gas.

\begin{figure}
 \centering 
		\gridline{\fig{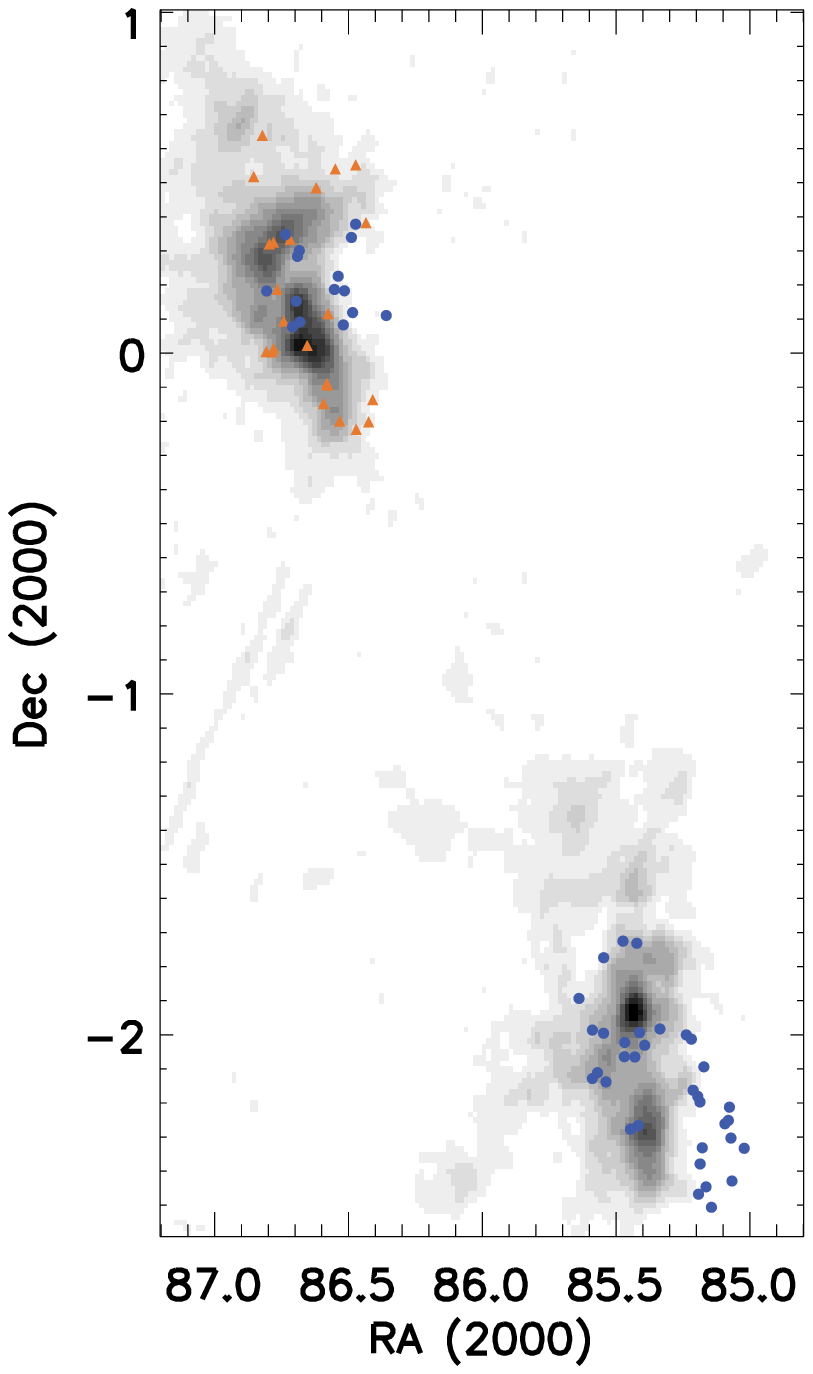}{0.1775\textwidth}{}
              \fig{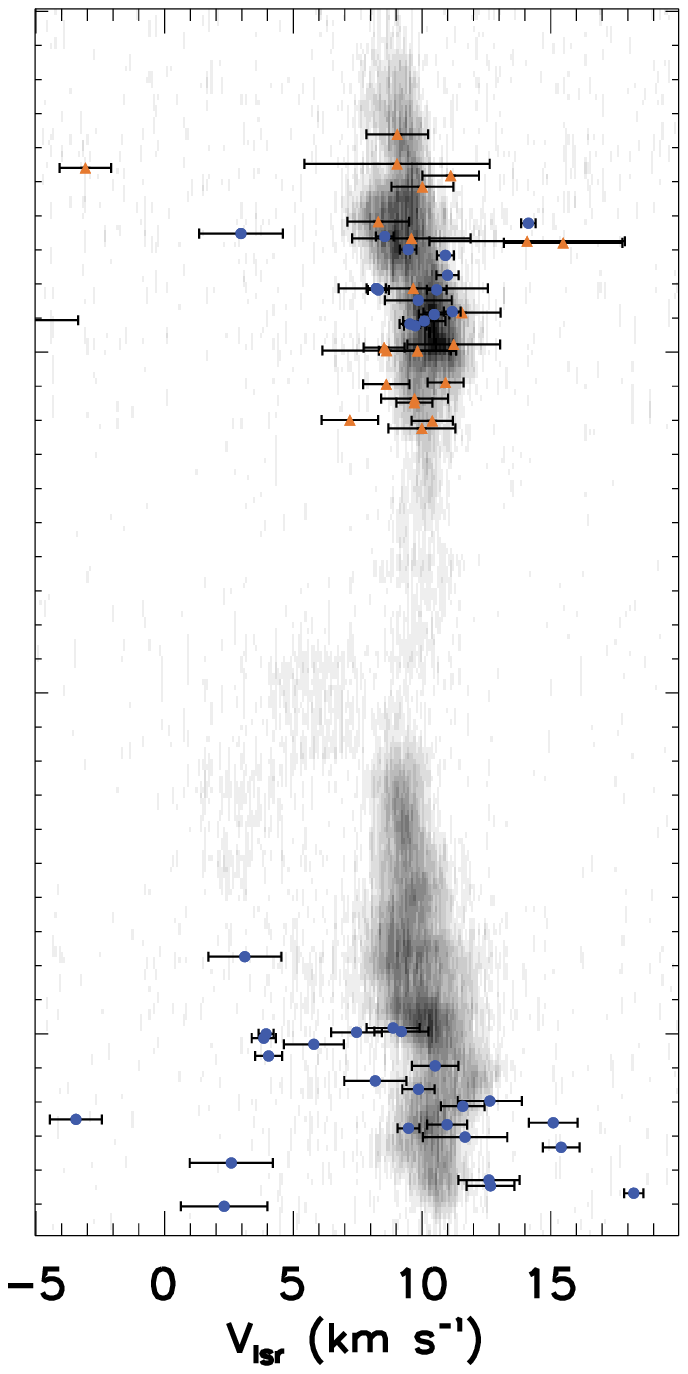}{0.15\textwidth}{}
              \fig{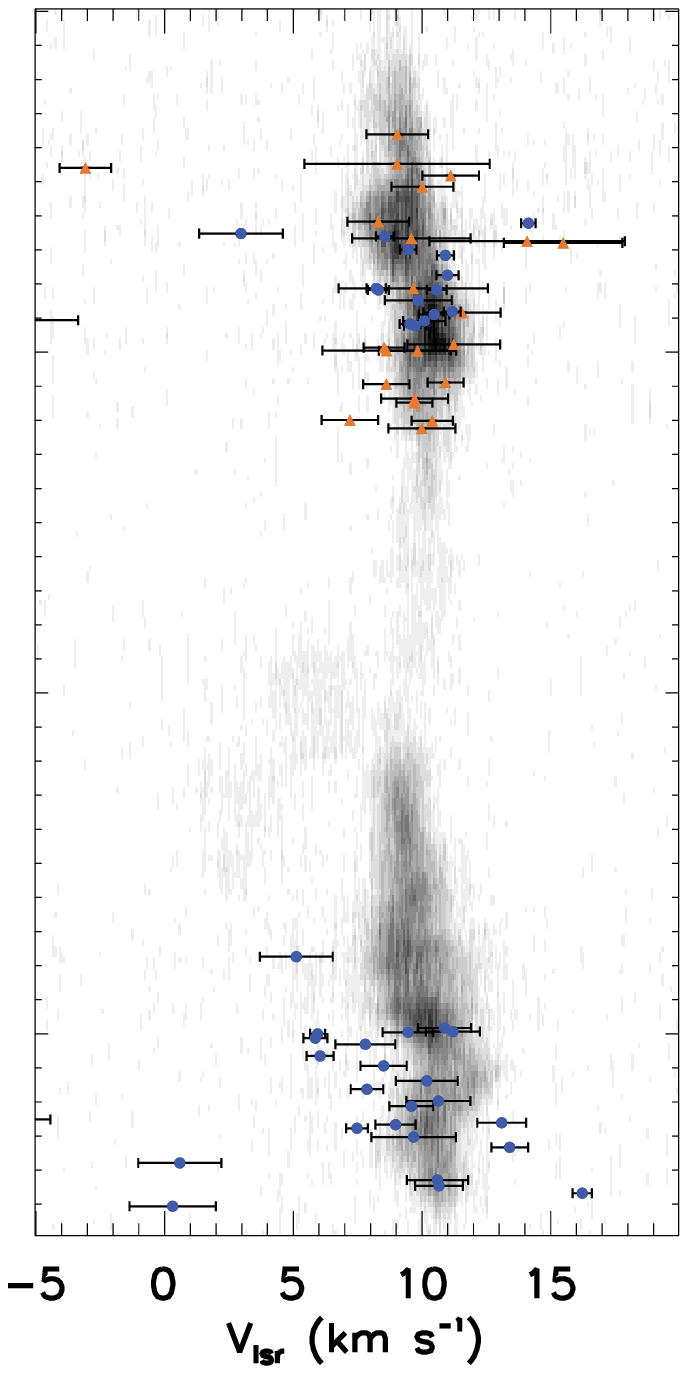}{0.15\textwidth}{}
        }
 \caption{Left: Map of the observed members of the NGC 2024 and NGC 2068 clusters, plotted over $^{13}$CO map from \citet{2015nishimura}. Blue circles are the sources from this work, orange triangles are the sources from FM08 survey. Middle: position-radial velocity diagram, summed in right ascention, using raw RVs. Right: same as before, but with the RVs of the two fields of NGC 2024 corrected for the offset. \label{fig:rv} }
\end{figure}

The systematic offset is much better calibrated in NGC 2068. After it is taken into the account, there does not appear to be any peculiar RV structure in NGC 2068. This is similar to what has been observed by FM08. We compare the RVs of the sources to the of $^{13}$CO gas \citep{2015nishimura}. Two sources appear to have RV different from the molecular gas (Figure \ref{fig:rv}). RV 2762 has been detected both in 2015 and 2017, and it has been observed by FM08 (source ID 984). There is a big degree of scatter in RV between the measurements. However, since all the measurements are highly uncertain, this source cannot be identified as a RV variable using the criterion from Paper I. Similarly, RV 2764 has been detected in all epochs (FM08 228). All the measurements are consistent with each other within the uncertainties (typically $\sim$0.5 km s$^{-1}$), all $\sim$5 km$^{-1}$ redshifted relative to the gas. While we cannot rule out that this source is  RV variable due to the sparsity of measurements, it is more likely that it has been ejected from the cluster.

NGC 2024, on the other hand, is significantly less clear-cut in terms of its RV structure. There appears to be a gradient across the cluster in the raw RVs of the members of the cluster, completely decoupled from the kinematics of the molecular gas. However as the observations of this region are split into two fields, it is possible that there is a systematic offset in both of these fields that is applied in the opposite direction from each other. The magnitude of the offset suggested by the telluric lines in both cases is $\sim$2 km s$^{-1}$. After this offset is removed from the data, the agreement between the kinematics of the molecular gas and stars becomes better; however there is a substantial population of stars blueshifted to the gas, and a few sources (mostly located off-cloud) remain to be somewhat redshifted. Alternative offset values cannot help to improve the agreement in the RVs further.

As only one epoch of observations is currently available for all of these sources, it is possible that several of these sources are spectroscopic binaries, therefore their measured RV is different from the true RV of the system. Paper I estimated that only 5--6\% of stars in the ONC and in NGC 2264 can be identified as spectroscopic binaries with the spectral resolution comparable to what is used in this survey. If this fraction is similar in the Orion B, we should expect only $\sim$3--4 spectroscopic binaries in the entire survey. While the multiplicity function does vary between different star-forming regions, the difference between RV of the gas and that of the detected stars cannot be due to binarity.

An important consideration in these observations is that since in the optical regime it is only possible to probe sources with very low extinction (Figure \ref{fig:color}), we are not sensitive toward the deeply embedded objects that are located within the molecular clouds. Therefore the entirety of our sample consists of the sources that are located near the upper layer of the cloud; they are of some of the closest YSOs to the Sun in these regions. We estimate that in the fields covered by observations, there should be $\sim$220 Class 0/I and II YSOs toward NGC 2068, $\sim$475 toward NGC 2024, and $\sim$30 toward L1622 based on the density maps by \citet{2016Megeath}. The number of Class III sources in these regions is more difficult to estimate, although, based on the observed ratio of Class II and Class III sources, it is possible that that the total number of YSOs would be $\sim$1.5 times higher than the value estimated based on the density of only Class 0/I and II YSOs.

With this in mind, since we observe only a very small fraction of the total number of sources, we cannot rule out a possibility that the RV of the entire population of the YSOs of NGC 2024 does follow the gas, with the same mean velocity and velocity dispersion, and that the sources we observe only the tail end of the Gaussian distribution. This would imply that the RV distribution in the cluster has some structure, and that both clusters are expanding. Another possibility is that these objects were dynamically ejected from the cluster.

However, the detected members could also trace the larger population of stars, the RV of which is systematically different from the gas, and that could be a result of gas being accelerated due to the stellar feedback, whereas the stars remain at the original RV of the gas at the time of the formation. The signature of it could potentially be seen in NGC 2023/2024 where the low density gas appears to be red-shifted relative to the two significantly denser clumps. Finally, this could be a signature of cold collapse; recent hydrodynamic simulations of globally gravitationally collapsing clouds demonstrate a similar assymetry between the distributions of young stars and molecular gas can be produced naturally toward massive clusters \citep{2017kuznetsova}. Ultimately, high resolution infrared spectra (such as the ones APOGEE will be able to provide) would be needed to obtain RV of a significantly larger number of stars embedded within the gas to confirm the degree to which the stars follow the gas.

The observations of other massive clusters, such as the ONC and NGC 2264 \citep{2016dario,2016kounkela} did reveal that, while the majority of stars do follow the gas, they also contain a significant blueshifted populations. Unfortunately the cause of it is still not entirely clear. Most likely it is a combination of the aforementioned reasons, although previous APOGEE observations of the ONC did mostly rule out extinction as the cause of the discrepancy. Observations of the Orion B put these observations into the perspective as they adds the significant information about the spacial distribution of the YSOs as a function of depth of the clusters.

\section{Conclusion}

We obtained optical high resolution spectra of 295 stars toward the Orion B molecular cloud. Of these sources, 67 can be identified as members of the clusters associated with the cloud on the basis of presence of strong Li I absorption, which ca be used as an indicator of youth in stars later than K4. It is more difficult to distinguish the young stars with earlier spectral types from their more evolved counterparts; therefore there may be some bona fide members that we have not identified as such.

Despite their signatures of youth, we have identified 7 late type sources in the Orion B and in the ONC with somewhat depleted Li I, with $0.1<W_{Li I}<0.4$. These sources should not be affected by veiling. Typically stars will achieve this level of depletion at the ages of 5--10 Myr \citep{1998baraffe}, which is significantly older than the estimated age of the clusters which these stars inhabit, and these sources may not necessarily be members of the clusters. However, it is possible that the accretion processes have accelerated the processing of Li I in these sources \citep{2016baraffe}. We also identified a number of sources with variable and/or assymetric H$\alpha$ line. Six of these sources are CTTSs, however, two sources are clearly more evolved and may not be associated with the cloud.

We measured RVs for all the sources. We find that members of NGC 2068 and L1622 tend to have RVs similar to those of the molecular gas. On the other hand, NGC 2024 does show a sizable population of stars that are preferentially blueshifted. Similar kinematics have been observed toward other clusters, such as the ONC and NGC 2264 (Paper I). It is still unclear what is the cause of it. Some possibilities include (a) sample bias due to e.g. extinction which prevents us from observing better agreement in RVs, (b) acceleration of the gas through stellar feedback, (c), observing older foreground population of stars that has dissipated their molecular gas and (d) a dynamical signature of cold collapse in the vicinity of a massive cluster.

Soon to be released distance and proper motion solutions by \textit{Gaia} DR2 will further contribute to the interpretation of the stellar dynamics of these clusters. As it is an optical telescope, it will not peer into the depth of the clusters, but it will be able to constrain the dynamics of the other two dimensions of motion of the stars presented in this paper. Additionally, APOGEE observations as part of the IN-SYNC program for the Orion B molecular cloud will allow RV measurements of much more embedded sources. Together, they will be instrumental in distinguishing between possibilities that may explain kinematic signatures observed in this paper.

\acknowledgments
The authors would like to thank the LCO operators and staff for their help on the observing run. This work was supported in part by the University of Michigan.

\software{TOPCAT \citep{topcat}, IRAF \citep{iraf1,iraf2}, RVSAO \citep{rvsao}}

\bibliographystyle{aasjournal.bst}

\input{m2fsfields.tex}
\input{li.tex}
\end{document}

%% file: m2fsfields.tex
\begin{deluxetable*}{cccccc}
\tabletypesize{\scriptsize}
\tablewidth{0pt}
\tablecaption{Dates and configurations of the M2FS observations.\label{tab:m2fsfield}}
\tablehead{
\colhead{Field} &\colhead{Date} & \colhead{R.A.} & \colhead{Dec.} & \colhead{Exposure time} & \colhead{Filter}\\
\colhead{ID} &\colhead{(UT)} & \colhead{(J2000)} & \colhead{(J2000)} & \colhead{(\#$\times$seconds)} & \colhead{}
}
\startdata
NGC 2024 & 2014 Dec 11 & 05:41:48.0 & $-$01:57:18 & 3$\times$1200 & Li I + H$\alpha$ \\
NGC 2023 & 2015 Feb 03 & 05:41:02.7 & $-$02:18:18 & 3$\times$1800 & Li I + H$\alpha$ \\
NGC 2068 & 2015 Feb 28 & 05:46:21.6 & 00:10:23 & 3$\times$1800 & Li I + H$\alpha$ \\
L1622 & 2015 Feb 27 & 05:54:20.1 & 01:42:57 & 3$\times$1800 & Li I + H$\alpha$\\
NGC 2068 & 2017 Mar 04 & 05:46:21.6 & 00:10:23 & 3$\times$720 & Li I + H$\alpha$ \\
NGC 2068 & 2017 Mar 04 & 05:46:21.6 & 00:10:23 & 3$\times$900 & Mg I \\
\enddata
\end{deluxetable*}

%% file: li.tex
\begin{deluxetable*}{ccccccccccccccc}
\tabletypesize{\scriptsize}
\tablewidth{0pt}
\tablecaption{Sources that were surveyed for the presence of Li I. \label{tab:li}}
\tablehead{
\colhead{RV} &\colhead{R.A.} &\colhead{Dec.} & \colhead{$v$} & \colhead{$\sigma$} & \colhead{$R$} & \colhead{$W_{Li}$}& \colhead{$W_{H_\alpha}$\tablenotemark{a}} & \colhead{$V_{H_\alpha}$} & \colhead{YSO} & \colhead{RR?\tablenotemark{b}} & \colhead{Temp} &  \colhead{Date} & \colhead{Order\tablenotemark{c}} & \colhead{Field}\\
\colhead{$\#$} &\colhead{(J2000)} &\colhead{(J2000)} & \colhead{km s$^{-1}$} & \colhead{km s$^{-1}$} & \colhead{} & \colhead{\AA}& \colhead{\AA}& \colhead{km s$^{-1}$}& \colhead{}& \colhead{}& \colhead{K.}& \colhead{JD} & \colhead{} & \colhead{}
}
\startdata
2533  & 05:40:58.97 & -02:32:52.0 & -16.39 &  0.87 &  10.51 & --- & --- & --- &  & y & 5750 & 2457084.5 & Li+H$\alpha$ & 2023 \\
2534  & 05:40:53.40 & -02:32:27.6 & -38.84 &  0.65 &  12.50 & --- &    1.35 & --- &  & y & 5500 & 2457084.5 & Li+H$\alpha$ & 2023 \\
2535  & 05:41:20.87 & -02:32:06.6 &  41.00 &  0.54 &  18.91 & --- &    1.24 & --- &  & y & 6000 & 2457084.5 & Li & 2023 \\
2536  & 05:41:12.08 & -02:31:21.9 & -31.40 &  0.96 &   7.12 & --- &    1.35 & --- &  & y & 5750 & 2457084.5 & Li+H$\alpha$ & 2023 \\
2537  & 05:41:05.44 & -02:31:06.4 &  91.65 &  0.36 &  21.98 & --- &    1.70 & --- &  & y & 5250 & 2457084.5 & Li+H$\alpha$ & 2023 \\
2538  & 05:40:47.38 & -02:30:52.9 &  32.13 &  1.19 &   6.35 & --- & --- & --- &  & y & 6250 & 2457084.5 & Li+H$\alpha$ & 2023 \\
2539  & 05:40:44.56 & -02:30:36.7 &  74.72 &  0.41 &  18.89 & --- &    2.44 & --- &  & y & 5750 & 2457084.5 & Li+H$\alpha$ & 2023 \\
2540  & 05:40:34.50 & -02:30:22.3 &  17.97 &  1.68 &  10.21 & 0.60 &   -1.54 & 289 & YSO (Class II)\tablenotemark{d} & y & 4000 & 2457084.5 & Li+H$\alpha$ & 2023 \\
\enddata
\tablenotetext{a}{Negative values show that the line appeared in emission. Most lines were contaminated by the nebular emission lines; this emission has been masked}
\tablenotetext{b}{Whether or not the source was processed with the rapid rotator parameters during cross-correlation (see Paper I for full discussion)}
\tablenotetext{c}{Order used for cross-correlation.}
\tablenotetext{d}{Not observed with \textit{Spitzer}, classified based on their H$\alpha$.}
\tablenotetext{}{Full version of the table will be available in the online text.}
\end{deluxetable*}